\documentstyle[epsfig,12pt]{article}

\newcommand{\bce}{\begin{center}}
\newcommand{\ece}{\end{center}}
\newcommand{\be}{\begin{equation}}
\newcommand{\ee}{\end{equation}}
\newcommand{\bea}{\begin{eqnarray}}

\newcommand{\eea}{\end{eqnarray}}

\newcommand{\ba}{\begin{array}}
\newcommand{\ea}{\end{array}}

\newcommand{\bkappa}{\mbox{\boldmath $\kappa$}}

\newcommand{\br}{{\bf r}}

\newcommand{\bk}{{\bf k}}

\def\Qb{\overline{Q}}

\def\lsim{\mathrel{\rlap{\lower4pt\hbox{\hskip1pt$\sim$}}
    \raise1pt\hbox{$<$}}}         
\def\gsim{\mathrel{\rlap{\lower4pt\hbox{\hskip1pt$\sim$}}
    \raise1pt\hbox{$>$}}}         

\def\Pom{{\bf I\!P}}

\def\Pom{{\bf I\!P}}

\def\lsim{\mathrel{\rlap{\lower4pt\hbox{\hskip1pt$\sim$}}
    \raise1pt\hbox{$<$}}}         

\def\gsim{\mathrel{\rlap{\lower4pt\hbox{\hskip1pt$\sim$}}
    \raise1pt\hbox{$>$}}}         

\date{}

\title {Anatomy of the differential gluon structure function of the proton from 
the experimental data on $F_{2p}(x,Q^{2})$}

\author{I.P. Ivanov$^{1,2,3}$\thanks{E-mail: i.ivanov@fz-juelich.de},
N.N.Nikolaev$^{1,4}$\thanks{E-mail: n.nikolaev@fz-juelich.de}\\
\makebox[8cm][c]{\normalsize
$^1$ IKP(Theorie), Forschungszentrum J\"ulich, Germany}\\
\makebox[8cm][c]{\normalsize
$^2$  Institute of Mathematics, Novosibirsk, Russia}\\
\makebox[8cm][c]{\normalsize
$^3$  Novosibirsk State University, Novosibirsk, Russia}\\
\makebox[8cm][c]{\normalsize
$^4$L. D. Landau Institute for Theoretical Physics, Moscow, Russia}}

\begin{document}
\maketitle
\vspace{-9cm}
\makebox[\textwidth][r]{\large\bf FZJ-IKP(Th)-2000/08}
\vspace{8cm}

\begin{abstract}
The use of the differential gluon structure function of the proton 
${\cal F}(x,Q^{2})$ introduced by Fadin, Kuraev and Lipatov in 1975
is called upon in many applications of small-$x$ QCD.  We report here
the first determination of ${\cal F}(x,Q^{2})$ from the experimental
data on the small-$x$ proton structure function $F_{2p}(x,Q^{2})$. 
We give convenient parameterizations for ${\cal F}(x,Q^{2})$ based partly
on the available DGLAP evolution fits (GRV, CTEQ \& MRS) to parton
distribution functions and on realistic extrapolations into soft region.
We discuss an impact of soft gluons on various observables. 
The $x$-dependence of the so-determined ${\cal F}(x,Q^{2})$  
varies strongly with $Q^2$ and does not exhibit simple Regge properties.
None the less the hard-to-soft diffusion is found to give rise to
a viable approximation of the proton structure function $F_{2p}(x,Q^2)$ 
by the soft and hard Regge components with intercepts 
$\Delta_{soft}=0$ and $\Delta_{hard}\sim 0.4$.
  
\end{abstract}
 

\section{Introduction: Why unintegrated gluon structure functions?}

The familiar objects from Gribov-Lipatov-Dokshitzer-Altarelli-Parisi
(DGLAP) evolution description of deep inelastic scattering (DIS) are
quark, antiquark and gluon distribution functions $q_{i}(x,Q^{2}),
\bar{q}(x,Q^{2}),g(x,Q^{2})$ (hereafter $x,Q^{2}$ are the standard 
DIS variables). At small $x$ they describe the integral flux of partons with 
the lightcone momentum $x$ in units of the target momentum and 
transverse momentum squared $\leq Q^{2}$ and form 
the basis of the highly sophisticated description of hard scattering 
processes in terms of collinear partons \cite{DGLAP}. On the other hand, at very 
small $x$ the object of the Balitskii-Fadin-Kuraev-Lipatov evolution 
equation is the differential gluon structure 
function (DGSF) of the target \cite{FKL}
\be
{\cal F}(x,Q^{2})={\partial G(x,Q^{2})\over \partial \log Q^{2}}
\label{eq:1.1}
\ee
(evidently the related unintegrated distributions can be defined also 
for charged partons). For instance, it is precisely DGSF of the target
proton which emerges in the familiar color dipole picture of inclusive 
DIS at small $x$ \cite{NZ94} and diffractive DIS into dijets 
\cite{NZsplit}. Another familiar example is the QCD 
calculation of helicity amplitudes of diffractive DIS into
continuum \cite{NPZLT,Twist4} and production of vector 
mesons \cite{Vmeson,JETPVM}. DGSF's are custom-tailored 
for QCD treatment of hard processes,
when one needs to keep track of the transverse momentum of gluons
neglected in the standard collinear approximation \cite{PomKperp}.

In the past two decades DGLAP phenomenology of DIS has become a big industry 
and several groups --- notably GRV \cite{GRV}, CTEQ \cite{CTEQ} \& MRS 
\cite{MRS} and some other \cite{BaroneF2} --- keep continuously incorporating 
new experimental data and providing the high energy community 
with updates of the parton distribution functions supplemented with the 
interpolation routines facilitating practical applications. On the other hand,
there are several pertinent issues --- the onset of the purely perturbative
QCD treatment of DIS and the impact of soft mechanisms of photoabsorption
on the proton structure function in the region of large $Q^{2}$ being 
top ones on the list --- which can not be answered within the DGLAP approach 
itself because DGLAP evolution is obviously hampered at moderate to
small $Q^{2}$. The related issue is to what extent the soft mechanisms of
photoabsorption can bias the $Q^{2}$ dependence of the proton structure
function and, consequently, determination of the gluon density from
scaling violations. We recall here the recent dispute 
\cite{dF2dLogQ2} on the applicability 
of the DGLAP analysis at $Q^{2} \lsim $ 2--4 GeV$^{2}$ triggered by the 
so-called Caldwell's plot \cite{Caldwell}. Arguably the 
$\bkappa$-factorization formalism of 
DGSF in which the interesting observables are expanded in interactions 
of gluons of transverse momentum $\bkappa$  changing 
from soft to hard is better suited to look into the issue of soft-hard
interface. At last but not the least, neglecting the transverse momentum
$\bkappa$ of gluons is a questionable approximation in evaluation of
production cross sections of jets or hadrons with large transverse 
momentum. It is distressing, then, that convenient parameterizations 
of DGSF are not yet available in the literature.

In this communication we report a simple phenomenological
determination of the DGSF of the proton at small $x$.  We analyze $x$
and $Q^{2}$ dependence of the proton structure function
$F_{2p}(x,Q^{2})$ in the framework of the $\bkappa$-factorization
approach, which is closely related to the color dipole factorization.
In the formulation of our Ansatz for ${\cal F}(x,\bkappa^2)$ we take
advantage of large body of the early work on color dipole
factorization \cite{NZ94,NZHERA,BFKLRegge} and follow a very pragmatic
strategy first applied in \cite{NPZLT,Twist4}: (i) for hard gluons
with large $\bkappa$ we make as much use as possible of the existing
DGLAP parameterizations of $G(x,\bkappa^{2})$, (ii) for the
extrapolation of hard gluon densities to small $\bkappa^{2}$ we use an
Ansatz \cite{NZsplit} which correctly describes the color gauge
invariance constraints on radiation of soft perturbative gluons by
colour singlet targets, (iii) as suggested by color dipole
phenomenology, we supplement the density of gluons with small
$\bkappa^{2}$ by nonperturbative soft component, (iv) as
suggested by the soft-hard diffusion inherent to the BFKL evolution,
we allow for propagation of the predominantly hard-interaction driven
small-$x$ rise of DGSF into the soft region invoking plausible
soft-to-hard interpolations. The last two components of DGSF are
parameterized following the modern wisdom on the infrared freezing of
the QCD coupling and short propagation radius of perturbative gluons. Having
specified the infrared regularization, we can apply the resulting
${\cal F}(x,\bkappa^2)$ to evaluation of the photoabsorption cross section in the
whole range of small to hard $Q^{2}$.

The practical realization of the above strategy is expounded as follows:
The subject of section 2 is a pedagogical introduction into the concept of
DGSF on an example of Fermi-Weizs\"acker-Williams photons in 
QED. Taking the electrically neutral positronium as a target, we explain 
important constraints imposed by gauge invariance on DGSF at small
$\kappa^{2}$. In section 3 we present the $\bkappa$-factorization
approach, which is the basis of our analysis of small-$x$ DIS 
in terms of DGSF. We also comment 
on the connection between the standard DGLAP analysis of DIS and 
$\bkappa$-factorization and property of 
soft-to-hard and hard-to-soft diffusion 
inherent to $\bkappa$-factorization. In section 4 we formulate our Ansatz for 
DGSF. The results of determination of DGSF from the experimental data on the
proton structure function $F_{2p}(x,Q^2)$ and real photoabsorption
cross section are presented in section 5. In section 6 we discuss an 
anatomy of the so-determined DGSF in the momentum space and comment
on the interplay of soft and hard components in DGSF, integrated gluon 
SF and proton structure function $F_{2p}(x,Q^2)$. In section 7 we focus on 
effective intercepts of $x$-dependence,  and the systematics of
their change from  DGSF to integrated gluon SF to $F_{2p}(x,Q^2)$,
which illustrates nicely the gross features of soft-to-hard and 
hard-to-soft diffusion pertinent to BFKL physics. The subject of
section 8 is a comparison of integrated gluon distributions 
from $\bkappa$-factorization and conventional DGLAP analysis of
the proton structure function. As anticipated, the two distributions
diverge substantially at very small $x$ and small to moderate $Q^2$. 
In section 9 discuss in more detail how different observables --- the scaling 
violations $\partial F_{2p}(x,Q^2)/\partial \log Q^2$, the longitudinal
structure function $F_{L}(x,Q^2)$ and charm structure function 
$F_{2}^{c\bar{c}}(x.Q^{2})$ ---  probe the DGSF. In section 10 we summarize
our major findings.


\section{Differential density of gauge bosons: the QED example}

\begin{figure}[!htb]
   \centering
   \epsfig{file=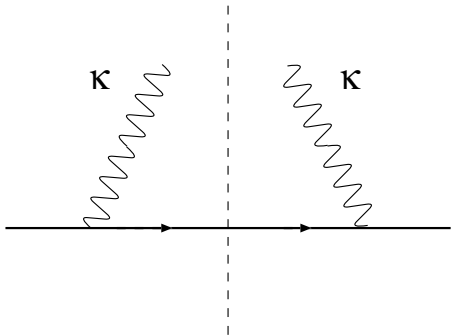,width=6cm}
   \caption{\em The Fermi-Weizs\"acker-Williams diagram for 
calculation of the flux of equivalent photons}
   \label{WWFermi}
\end{figure}

For the pedagogical introduction we recall the 
celebrated Fermi-Weizs\"acker-Williams approximation in QED, which is the
well known precursor of the parton model (for 
the review see \cite{Budnev}). Here 
high energy reactions in the  Coulomb field of a charged particle are treated 
as collisions with equivalent transversely polarized photons --- partons of 
the charged particle, fig.\ref{WWFermi}. 
The familiar flux of comoving equivalent transverse 
soft photons carrying a lightcone fraction $x_{\gamma} \ll 1$ of the momentum of a 
relativistic particle, let it be the electron, reads 
\be
dn^\gamma_e = {\alpha_{em} \over \pi }
{\bkappa^2 d\bkappa^2 \over (\bkappa^2 +\kappa_{z}^{2})^{2}} 
{dx_{\gamma} \over x_{\gamma}}
\approx {\alpha_{em} \over \pi }
{d\bkappa^2  \over \bkappa^2 } 
{dx_{\gamma} \over x_{\gamma}}\,,
\label{eq:2.1}
\ee
Here $\bkappa$ is photon transverse momentum and $\kappa_{z}=m_{e}x_{\gamma}$ 
is the photon longitudinal momentum in the electron Breit frame.
The origin of $\bkappa^2$ in the numerator is in the current conservation, 
i.e. gauge invariance. Then the unintegrated photon structure function 
of the electron is by definition
\be
{\cal F}_{\gamma}(x_{\gamma},\bkappa^2)
 = { \partial G_\gamma \over \partial \log \bkappa^2} =
x_{\gamma}{dn^\gamma_e \over dx_{\gamma} d\log\bkappa^{2}}
= {\alpha_{em} \over \pi}\left( {\bkappa^2 \over 
\bkappa^2 +\kappa_{z}^{2}}\right)^{2}
\,.
\label{eq:2.2}
\ee

\begin{figure}[!htb]
   \centering
   \epsfig{file=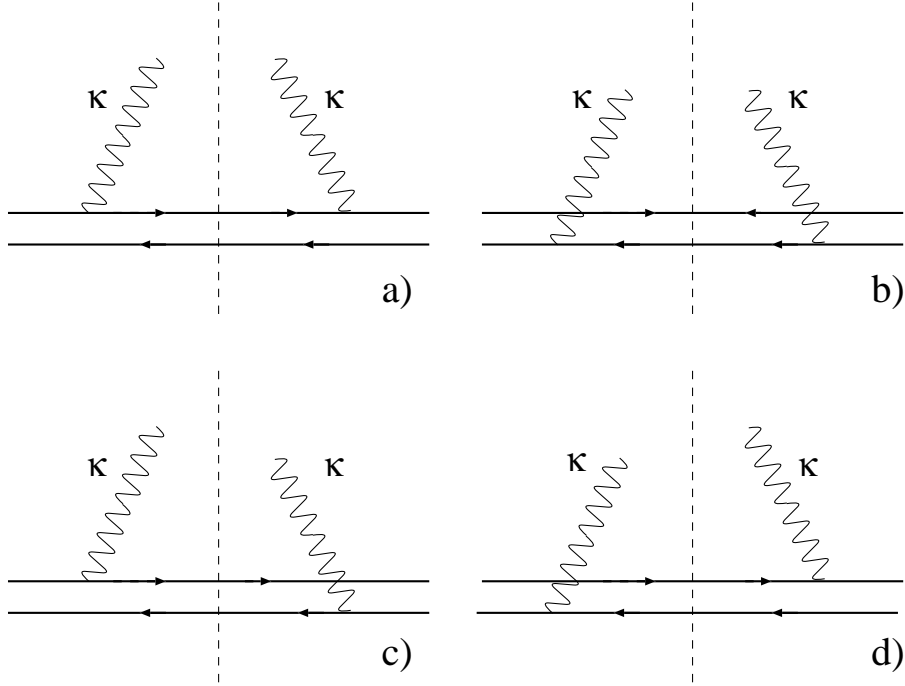,width=12cm}
   \caption{\em The Fermi-Weizs\"acker-Williams diagrams for 
calculation of the flux of equivalent photons in positronium.}
   \label{WWpositronium}
\end{figure}

If the relativistic particle is a positronium, fig.~\ref{WWpositronium},
destructive interference 
of electromagnetic fields of the electron and positron must be taken 
into account. Specifically, for soft photons with the wavelength 
$\lambda = {1\over \kappa} \gg a_{{\rm P}}$, where  $a_{{\rm P}}$ is the 
positronium Bohr radius, the electromagnetic fields of an electron
and positron cancel each other and the flux of photons vanishes,
whereas for $\lambda \ll a_{{\rm P}}$ the flux of photons will be twice
that for a single electron.
The above properties are quantified by the formula
\be
{\cal F}_{\gamma}^{{\rm P}}(x_{\gamma},\bkappa^2) = 
2{\alpha_{em} \over \pi} \left( {\bkappa^2 \over 
\bkappa^2 +\kappa_{z}^{2}}\right)^{2} V(\kappa)\,,
\label{eq:2.3}
\ee
where the factor 2 is a number of charged particles in the positronium
and corresponds to the Feynman diagrams of fig.~\ref{WWpositronium}a,~\ref{WWpositronium}b. 
The vertex function $V(\kappa)$ is expressed in terms of the two-body 
formfactor of the positronium,
\be
V(\kappa) = 1 - F_2(\bkappa,-\bkappa) = 
1 - \langle{\rm P}|\exp(i\bkappa (\br_{-} - \br_{+}))|{\rm P}\rangle\,,
\label{eq:2.4}
\ee
where $\br_{-} - \br_{+}$ is the spatial separation of $e^+$ and $e^{-}$
in the positronium.
The two-body formfactor $F_{2}(\bkappa,-\bkappa)$ 
describes the destructive interference of electromagnetic fields of the 
electron and positron and corresponds to the Feynman diagrams 
of fig.~\ref{WWpositronium}c,~\ref{WWpositronium}d. 
It vanishes for large enough $\kappa \gsim a_{{\rm P}}^{-1}$, 
leaving us with $V(\kappa)=2$, whereas for soft gluons one has
\be 
V(\kappa) \propto \bkappa^{2}a_{{\rm P}}^{2}
\label{eq:2.5}
\ee
One can say that the law (\ref{eq:2.5}) is driven by electromagnetic gauge 
invariance, which guarantees that long wave photons decouple 
from the charge neutral system. 

Finally, recall that the derivation of the differential flux
of transverse polarized photons would equally hold if the 
photons were massive vector bosons interacting with the conserved current, 
the only change being in the propagator. For instance, for the charge
neutral source one finds
\be
{\cal F}_{V}^{{\rm P}}(x_{V},\bkappa^2) = 
{\alpha_{em} \over \pi} \left( {\bkappa^2 \over 
\bkappa^2 +m_{V}^{2}}\right)^{2} V(\kappa)\,.
\label{eq:2.6}
\ee
Recall that the massive vector fields are Yukawa-Debye screened with the screening
radius 
\be
R_{c}=m_{V}^{-1}\, .
\label{eq:2.7}
\ee
To the lowest in QED perturbation theory the two exchanged photons in
figs.\ref{WWFermi},~\ref{WWpositronium} do not interact 
and we shall often refer to (\ref{eq:2.6}) as
the Born approximation for the differential vector boson structure function.
One can regard (\ref{eq:2.6}) as a minimal model for soft $\kappa$ 
behavior of differential structure function for Yukawa-Debye 
screened vector bosons.


\section{The insight into the differential density of gluons}

\subsection{Modeling virtual photoabsorption in QCD}

The quantity which is measured in deep inelastic leptoproduction is the
total cross section of photoabsorption $\gamma^{*}_{\mu}p \to X$ summed over
all hadronic final states $X$, where $\mu,\nu=\pm 1,0$ are helicities of 
$(T)$ transverse and $(L)$ longitudinal virtual photons. One usually
starts with the imaginary part of the amplitude $A_{\mu\nu}$ 
of forward Compton scattering $\gamma_{\mu}^{*}p \to \gamma_{\nu}^{*}
p'$, which by optical theorem gives the total cross cross section 
of photoabsorption of virtual photons
\bea
\sigma_{T}^{\gamma^{*}p}(x_{bj},Q^2) = 
{1\over \sqrt{(W^{2}+Q^{2}-m_{p}^{2})^{2}+4Q^{2}m_{p}^{2}}}
{\rm Im} A_{\pm\pm}\, ,
\label{eq:3.1.1}
\eea
\be
\sigma_{L}^{\gamma^{*}p}(x_{bj},Q^2) = 
{1\over \sqrt{(W^{2}+Q^{2}-m_{p}^{2})^{2}+4Q^{2}m_{p}^{2}}}
{\rm Im} A_{00}\, ,
\label{eq:3.1.2}
\ee
where $W$ is the total energy in the $\gamma^{*}p$ {\it c.m.s.}, $m_{p}$ is
the proton mass, $Q^{2}$ is the virtuality of the photon
and  $x_{bj}=Q^{2}/(Q^{2}+W^{2} - m_{p}^{2})$ is the Bjorken variable.

\begin{figure}[!htb]
   \centering
   \epsfig{file=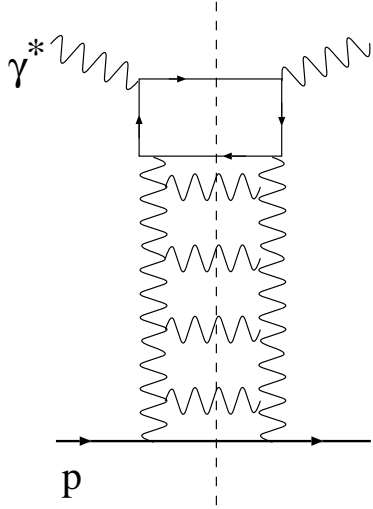,width=5cm}
   \caption{\em The pQCD modeling of DIS in terms of multiproduction
of parton final states.}
   \label{Multiproduction}
\end{figure}
In perturbative QCD (pQCD) one models virtual photoabsorption in terms of the 
multiple production of gluons, quarks and antiquarks 
(fig.~\ref{Multiproduction}). The experimental 
integration over the full phase space of hadronic states $X$ is substituted
in the pQCD calculation by integration over the whole phase space of QCD
partons
\be
\int |M_{{\cal X}}|^{2} d\tau_{{\cal X}} \Rightarrow \sum_{n} |M_{n}|^{2}
\prod\int_{0}^{1} {dx_{i} \over x_{i}} d^{2}{\bkappa}_{i}\, ,
\label{eq:3.1.3}
\ee
where the integration over the transverse momenta of partons goes over the
whole allowed region 
\be
0 \leq \kappa_{i}^{2} \leq {1\over 4}W^{2} = {Q^{2} (1-x) \over 4x}\, .  
\label{eq:3.1.4}
\ee
The core of the so-called DGLAP approximation \cite{DGLAP} is an observation that
at finite $x$ the dominant contribution to the multiparton production cross 
sections comes from a tiny part of the phase space
\bea
&1 \geq x_{1}  \geq x_{2} ... \geq  x_{n-1}  \geq  x_{n}  \geq  x &\, ,
 \nonumber\\
&0 \leq \kappa_{1}^{2} \ll \kappa_{2}^{2}...\ll\kappa_{n-1}^{2}
\ll k^{2} \ll  Q^2 &\,,\label{eq:3.1.5}
\eea   
in which the upper limit of integration over transverse momenta of partons is
much smaller than the kinematical limit (\ref{eq:3.1.4}). At very small $x$
this limitation of the transverse phase 
space becomes much too restrictive and the
DGLAP approximation is doomed to failure. 

\begin{figure}[!htb]
   \centering
   \epsfig{file=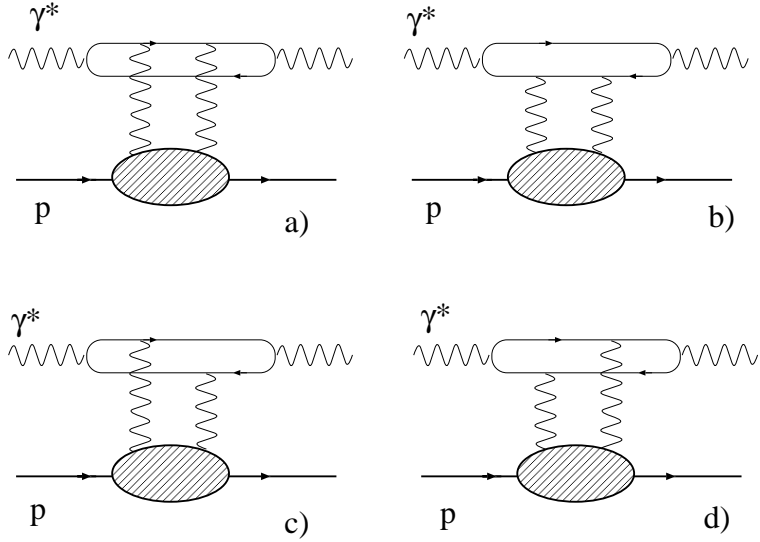,width=10cm}
   \caption{\em The $\bkappa$-factorization representation for
DIS at small $x$.}
   \label{KappaFactorization}
\end{figure}

Hereafter we focus on how lifting the restrictions on the transverse phase 
space changes our understanding of the gluon structure function of the
nucleon at very small $x$, that is, very large ${1\over x}$\,. In this kinematical
region the gluon density $g(x,Q^{2})$ is much higher than the density of 
charged partons $q(x,Q^{2}), \bar{q}(x,Q^{2})$. As Fadin, Kuraev and Lipatov
\cite{FKL} have shown, to the leading $\log {1\over x}$
(LL${1\over x}$) approximation the dominant contribution to 
photoabsorption comes  in this regime from multigluon 
final states of fig.~\ref{Multiproduction}; 
alternatively, to the LL${1\over x}$ splitting of gluons into gluons dominates 
the splitting of gluons into $q\bar{q}$ pairs. 
As a matter of fact, for the purposes of the present analysis
we do not need the full BFKL dynamics, in the 
$\bkappa$-factorization only the $q\bar{q}$ loop is treated explicitly 
to the  LL${1\over x}$ approximation. In this 
regime the Compton scattering can be viewed as an interaction of the nucleon with
the lightcone $q\bar{q}$  Fock states of the photon via the exchange by gluons,
fig.~\ref{KappaFactorization},  
and the Compton scattering amplitude takes the form
\be
A_{\nu\mu}=\Psi^{*}_{\nu,\lambda\bar{\lambda}}\otimes A_{q\bar{q}}\otimes
\Psi_{\mu,\lambda\bar{\lambda}}
\label{eq:3.1.6}
\ee 
Here $\Psi_{\mu,\lambda\bar{\lambda}}$ is the $Q^{2}$ and $q,\bar{q}$ helicity 
$\lambda,\bar{\lambda}$ dependent lightcone wave function of the photon and 
the  QCD pomeron exchange $q\bar{q}$-proton scattering kernel $A_{q\bar{q}}$  
does not depend on, and conserves exactly, the $q,\bar{q}$ helicities, 
summation over which is understood in (\ref{eq:3.1.6}).

The resummation of diagrams of fig.~\ref{Multiproduction} 
defines the unintegrated gluon structure function of the target,
which is represented in diagrams of fig.~\ref{KappaFactorization}
as the dashed blob. 
The calculation of the forward Compton scattering amplitudes 
(${\bf \Delta}=0$) is straightforward and gives the $\bkappa$-factorization
formulas for photoabsorption 
cross sections \cite{NZ91,NZglue} 
\bea
\sigma_{T}(x_{bj},Q^{2})=
{\alpha_{em}\over \pi}\sum_{f} e_{f}^{2}
\int_{0}^{1} dz \int d^{2}\bk \int
{ d^{2}\bkappa
\over \bkappa^{4}}\alpha_{S}(q^{2})
{\cal F}(x_{g},\bkappa^{2})
\nonumber\\
\left\{[z^{2}+(1-z)^{2}] 
\left({\bk \over  \bk^{2}+\varepsilon^{2}} - 
{\bk-\bkappa \over  (\bk-\bkappa)^{2}+\varepsilon^{2}}\right)^{2}\right.
\nonumber\\
\left.
+m_{f}^{2}
\left({1   \over  \bk^{2}+\varepsilon^{2}}-
{1 \over  (\bk-\bkappa)^{2}+\varepsilon^{2}}\right)^{2}\right\}
\label{eq:3.1.7}
\eea
\bea
\sigma_{L}(x_{Bj},Q^{2})=
{\alpha_{em}\over \pi}\sum_{f} e_{f}^{2}
\int_{0}^{1} dz \int d^{2}\bk \int{ d^{2}\bkappa
\over \bkappa^{4}}\alpha_{S}(q^{2})
{\cal F}(x_{g},\bkappa^{2}) 
\nonumber\\
4Q^{2}z^{2}(1-z)^{2}
\left({1   \over  \bk^{2}+\varepsilon^{2}}-
{1 \over  (\bk-\bkappa)^{2}+\varepsilon^{2}}\right)^{2}
\label{eq:3.1.8}
\eea
Here $m_{f}$ and $e_{f}$ are the mass and charge of the quark $f=u,d,s,c,b,..$,
\be
\varepsilon^2=z(1-z)Q^{2}+m_{f}^{2}\, ,
\label{eq:3.1.9}
\ee
the QCD running coupling $\alpha_{S}(q^{2})$ enters the integrand at the largest
relevant virtuality,
\be
q^{2}={\rm max }\{\varepsilon^{2}+\bk^{2},\bkappa^{2}\}\, ,
\label{eq:3.1.10}
\ee
and the density of gluons enters at
\be
x_{g}={Q^{2} +M_{t}^{2} \over W^{2}+Q^{2}}=
x_{bj}\left(1 +{M_{t}^{2} \over Q^{2}}\right) \,.
\label{eq:3.1.11}
\ee
Here $M_{t}$ is the transverse mass of the produced $q\bar{q}$
pair in the photon-gluon fusion $\gamma^{*}g\rightarrow q\bar{q}$:
\be
M_{t}^{2} = {m_{f}^{2}+\bk^{2} \over 1-z}+
{m_{f}^{2}+(\bk-\bkappa)^{2} \over z}   \, .
\label{eq:3.1.12}
\ee

For longitudinal photons only transitions $\gamma_{L} \to q_{\lambda} 
\bar{q}_{\bar{\lambda}}$ into states with ${\lambda}+\bar{\lambda}=0$ 
are allowed. In $\sigma_{T}$ the terms $\propto m_{f}^{2}$ 
are the contribution of states with ${\lambda}+\bar{\lambda}=\mu$, 
whereas the dominant contribution in the scaling regime 
of $Q^{2} \gg m_{f}^2$ comes from the transitions $\gamma_{T} 
\to q_{\lambda} \bar{q}_{\bar{\lambda}}$ into states ${\lambda}+\bar{\lambda}=0$,
when the helicity of the photon is transferred to the angular momentum of
the quark-antiquark pair. The corresponding transition amplitudes are
$\propto \bk,\bk\pm \bkappa$, for more discussion see \cite{Vmeson}.

No restrictions on the transverse momentum in the $q\bar{q}$ loop, 
$\bk$, and gluon momentum, $\bkappa$, are imposed in the representation 
(\ref{eq:3.1.7}), (\ref{eq:3.1.8}). 
This representation was contained essentially 
in the classic Fadin, Kuraev, Lipatov  papers \cite{FKL} of mid-70's, 
in the recent literature it is sometimes referred to as the 
$\bkappa$-factorization.

We note that Eqs. (\ref{eq:3.1.7}), (\ref{eq:3.1.8}) are for forward 
diagonal Compton
scattering, but similar representation in terms of the unintegrated gluons structure 
function holds also for the off-forward Compton scattering at finite momentum 
transfer ${\bf \Delta}$, off-diagonal Compton scattering 
when the virtualities of the initial and final state photons are different, 
$Q_{f}^{2}\neq Q_{i}^{2}$, including the timelike photons and vector mesons, 
$Q_{f}^{2}=-m_{V}^{2}$, in the final state. 

The photoabsorption cross sections define the dimensionless structure functions 
\be
F_{T,L}(x_{bj},Q^{2})={Q^{2}\over 4\pi^{2}\alpha_{em}}\sigma_{T,L}
\label{eq:3.1.13}
\ee
and $F_{2}=F_{T}+F_{L}$, which admit the familiar pQCD parton model interpretation
\be
F_{T}(x_{bj},Q^{2})= \sum_{f=u,d,s,c,b,..} e_{f}^{2}
[q_{f}(x_{bj},Q^{2})+\bar{q}_{f}(x_{bj},Q^{2})]\, ,
\label{eq:3.1.14}
\ee
where $q_{f}(x_{bj},Q^{2}), \bar{q}_{f}(x_{bj},Q^{2})$ are the integrated 
density of quarks and antiquarks carrying the fraction $x_{bj}$ 
of the lightcone momentum of the target and with 
transverse momenta $\leq Q$. Hereafter  we suppress the
subscript $"bj"$.


\subsection{Where $\bkappa_{\perp}$-factorization meets DGLAP factorization}

Recall the familiar DGLAP equation \cite{DGLAP} for scaling violations at small $x$,
\be
{d F_{2}(x,Q^{2}) \over d\log Q^{2}}= \sum_{f} e_{f}^{2}
{\alpha_{S}(Q^{2}) \over 2\pi} \int_{x}^{1} dy [y^2+(1-y)^{2}]G({x\over y},Q^{2})
\approx {\alpha_{S}(Q^{2}) \over 3\pi} G(2x,Q^{2})\sum_{f} e_{f}^{2}\, ,
\label{eq:3.2.1}
\ee
where for the sake of simplicity we only consider light flavours. Upon integration 
we find 
\be
F_{2}(x,Q^{2})\approx \sum_{f} e_{f}^{2}\int_{0}^{Q^{2}} {d\Qb^{2} \over \Qb^{2}}
{\alpha_{S}(\Qb^{2}) \over 3\pi}
G(2x,\Qb^{2})\, .
\label{eq:3.2.2}
\ee
In order to see the correspondence between the $\bkappa_{\perp}$-factorization and 
DGLAP factorization
it is instructive to follow the derivation of (\ref{eq:3.2.2}) from the 
$\bkappa_{\perp}$-representation (\ref{eq:3.1.7}).

First, separate the 
$\kappa^{2}$-integration in (\ref{eq:3.1.7}) into the DGLAP part
of the gluon phase space 
$\kappa^{2} \lsim \Qb^{2}=\epsilon^2 +k^2$ and beyond-DGLAP region 
$\kappa^{2} \gsim \Qb^{2}$. One readily finds
\be
\left({\bk \over  \bk^{2}+\varepsilon^{2}} - 
{\bk-\bkappa \over  (\bk-\bkappa)^{2}+\varepsilon^{2}}\right)^{2}
=
\left\{\begin{array}{lcr}
\left({2z^2(1-z)^2Q^4 \over \Qb^8}-{2z(1-z)Q^2 \over \Qb^{6}} +
{1\over \Qb^{4}}\right)\bkappa^2
& \mbox{if}& \bkappa^2 \ll \Qb^{2} \\[2mm]
\left({1\over \Qb^{2}}-{z(1-z)Q^2 \over \Qb^{4}}\right),& 
\mbox{if}& \bkappa^{2} \gsim \Qb^{2}
\end{array}\right.
\label{eq:3.2.3}
\ee

Consider first the contribution from the DGLAP part of the phase space
$\kappa^{2} \lsim \Qb^{2}$.
Notice that because of the factor $\bkappa^{2}$ in (\ref{eq:3.2.3}), 
the straightforward $\bkappa^2$ integration of the DGLAP component 
yields $G(x_{g},\Qb^{2})$ and $\Qb^2$ is precisely the pQCD hard scale
for the gluonic transverse momentum scale:
\bea
\int^{\Qb^2}_{0}
{ d\bkappa^2
\over \bkappa^{2}}\alpha_{S}(q^{2})
{\cal F}(x_{g},\bkappa^{2}) 
\left({\bk \over  \bk^{2}+\varepsilon^{2}} - 
{\bk-\bkappa \over  (\bk-\bkappa)^{2}+\varepsilon^{2}}\right)^{2}
\nonumber\\
=\left({2z^2(1-z)^2Q^4 \over \Qb^8}-{2z(1-z)Q^2 \over \Qb^{6}} +
{1\over \Qb^{4}}\right)G(x_{g},\Qb^{2})
\label{eq:3.2.4}
\eea
The contribution from the beyond-DGLAP region of the phase space 
can be evaluated as
\bea
\int_{\Qb^2}^{\infty}
{ d\bkappa^2
\over \bkappa^{4}}\alpha_{S}(q^{2})
{\cal F}(x_{g},\bkappa^{2})\left({1\over \Qb^{2}}-{z(1-z)Q^2 
\over \Qb^{4}}\right) 
=\left({1\over \Qb^{4}}-{z(1-z)Q^2 \over \Qb^{6}}\right)
{\cal F}(x_{g},\Qb^{2})I(x_{g},\Qb^2)\nonumber\\
=
\left({2z^2(1-z)^2Q^4 \over \Qb^8}-{2z(1-z)Q^2 \over \Qb^{6}} +
{1\over \Qb^{4}}\right){\cal F}(x_{g},\Qb^{2})\log C_{2}(x_{g},\Qb^2,z)\, .
\label{eq:3.2.5}
\eea
The latter form of (\ref{eq:3.2.5}) allows to conveniently combine (\ref{eq:3.2.4})
and (\ref{eq:3.2.5}) rescaling the hard scale in the GSF
\be
G(x_{g},\Qb^{2})+{\cal F}(x_{g},\Qb^{2})\log C_2(x_{g},\Qb^2,z)
=G(x_{g},C_{2}(x_{g},\Qb^2,z)\Qb^2)\, .
\label{eq:3.2.6}
\ee
Here the exact value of $I(x_{g},\Qb^2) \ge 1$ depends on 
the rate of the $\bkappa^2$-rise of ${\cal F}(x_{g},\bkappa^{2})$.
At small $x_{g}$ and small to moderate $\Qb^{2}$ one finds  
$I(x_{g},\Qb^2)$  substantially larger than 1 and 
$C_{2}(x_{g},\Qb^2,z)\gg 1$, see more discussion below in section 9. 

Now change from $d\bk^{2}$ integration to $d\Qb^2$ and again split the 
$z$,$Q^2$ integration into the DGLAP part of the phase space 
$\Qb^{2} \ll {1 \over 4}Q^2$, where 
either $z < {\Qb^2 \over Q^{2}}$ or $1-z < {\Qb^2 \over Q^{2}}$,  and 
the beyond-DGLAP region  $\Qb^{2} \gsim {1\over 4}Q^2$, where $0< z < 1$. 
As a result one finds 
\bea
\int dz [z^2+(1-z)^2]\left({2z^2(1-z)^2Q^4 \over \Qb^8}-{2z(1-z)Q^2 \over \Qb^{6}} +
{1\over \Qb^{4}}\right)\nonumber\\
 =
\left\{\begin{array}{lcr}
{4 \over 3\Qb^2 Q^2}, 
& \mbox{if}&  \Qb^{2}\ll {1\over 4}Q^{2} \\[2mm]
\left(2A_{2}{Q^4\over \Qb^{8}}-2A_{1}{Q^{2} \over \Qb^6} +A_{0}{1\over \Qb^{4}}\right),& 
\mbox{if}& \Qb^{2} \gsim {1\over 4}Q^{2}
\end{array}\right.
\label{eq:3.2.7}
\eea
where
\be
A_{m}=\int_{0}^{1} dz [z^2+(1-z)^2]z^m(1-z)^m
\label{eq:3.2.8}
\ee
Let $\overline{C}_2$ be $C_{2}(x_{g},\Qb^2,z)$ at a mean point. Notice also
that $M_{t}^{2} \sim Q^{2}$, so that $x_{g}\sim 2x$. Then the 
contribution from the DGLAP phase space of $\Qb^2$ can be cast in precisely
the form (\ref{eq:3.2.2})
\be
\left. F_{2}(x,Q^{2})\right|_{DGLAP}\approx \sum_{f} 
e_{f}^{2}\int_{0}^{{\overline{C_2}\over 4}Q^{2}} {d\Qb^{2} \over \Qb^{2}}
{\alpha_{S}(\Qb^{2}) \over 3\pi}
G(2x,\Qb^{2})\, .
\label{eq:3.2.9}
\ee

The beyond-DGLAP region of the phase space gives the extra contribution of
the form
\bea
\left. \Delta F_{2}(x,Q^{2})\right|_{non-DGLAP}\sim
\sum_{f} e_{f}^{2}
{\alpha_{S}(Q^2) \over 3\pi}
\int_{Q^{2}}^{\infty} 
{d\Qb^{2} \over \Qb^{2}}{Q^2 \over \Qb^2}G(2x,\Qb^{2}) \nonumber \\
\sim 
\sum_{f} e_{f}^{2}
{\alpha_{S}(Q^2) \over 3\pi}
G(2x,Q^{2})  
\, .
\label{eq:3.2.10}
\eea
Eqs.(\ref{eq:3.2.9}) and (\ref{eq:3.2.10}) immediately reveal
the phenomenological consequences of lifting the DGLAP restrictions
in the transverse momenta integration.
Indeed, the DGLAP approach respects the following strict inequalities
\be
\bkappa^2 \ll \bk^2 \quad \mbox{and} \quad 
\bk^2 \ll Q^2\,.
\ee
As we just saw, removing the first limitation effectively shifted
the upper limit in the $\Qb^2$ integral to
${{\overline C_2} \over 4} Q^2 \not = Q^2$, while lifting the second constraint
led to an additional, purely non-DGLAP contribution.
Although both of these corrections lack one leading log-$Q^2$ factor
they are numerically substantial.
As a matter of fact, in section 9 we show that ${\overline C_2 }\approx 8$.

The above analysis suggests that the DGLAP and $\bkappa$-factorization
schemes converge logarithmically at large $Q^{2}$. 
However, in order to reproduce the 
result (\ref{eq:3.2.9}) and (\ref{eq:3.2.10}) for the full phase space 
by the conventional DGLAP contribution (\ref{eq:3.2.2}) from the restricted 
phase space (\ref{eq:3.1.5}) one has to ask for DGLAP gluon density 
$G_{pt}(x,Q^{2})$ larger than the integrated GSF 
in the $\bkappa$-factorization scheme and the
difference may be quite substantial in the domain of strong scaling
violations.


\subsection{The different evolution paths: soft-to-hard diffusion and 
vice versa}

\begin{figure}[!htb]
   \centering
  \epsfig{file=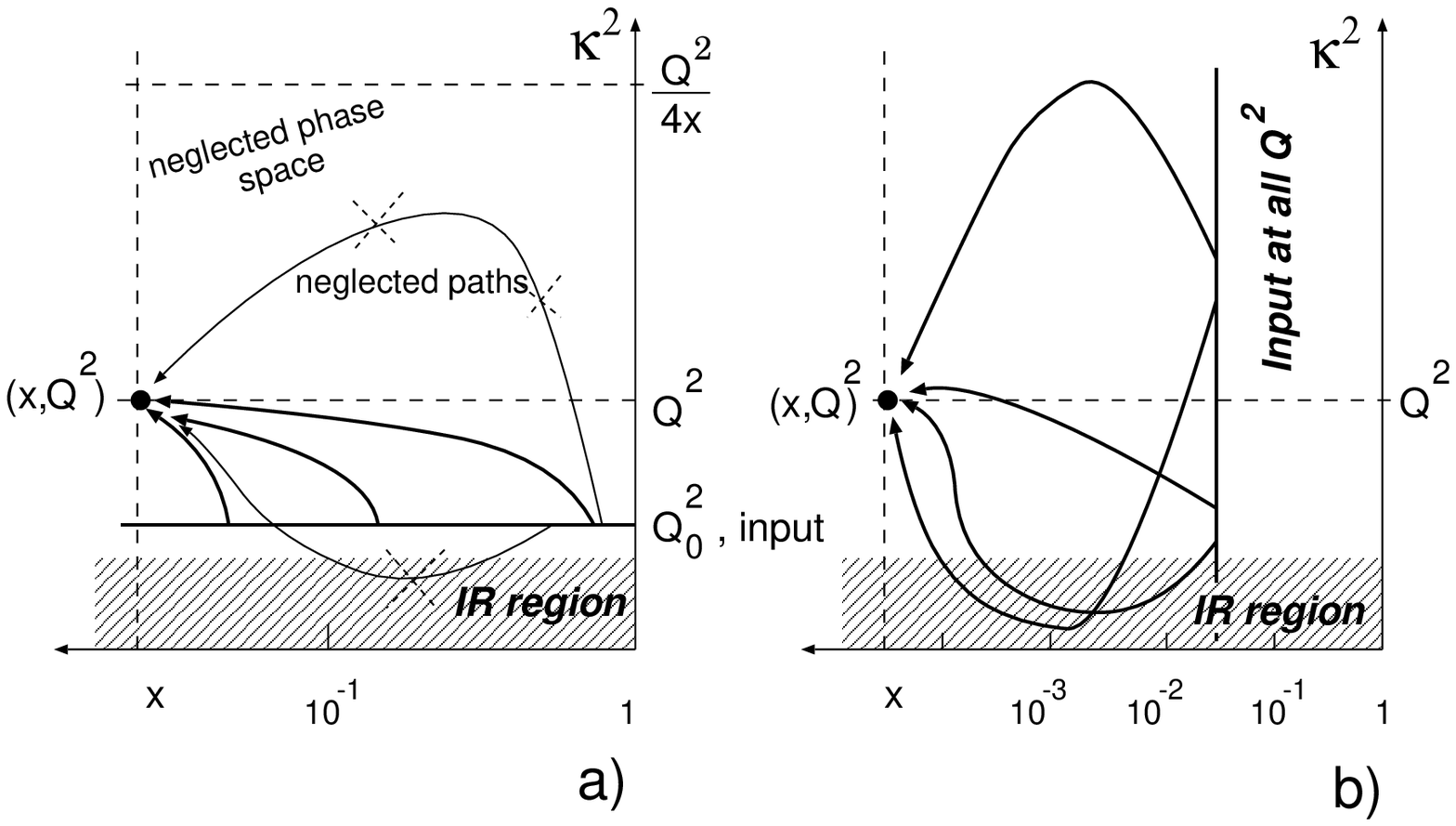,width=16cm}
   \caption{\em The Huygens principle for $Q^{2},x$ 
evolution of DIS structure
functions with {\rm (a)} DGLAP restricted transverse phase space and {\rm (b)}
for the BFKL
$x$ evolution without restrictions on the transverse phase space
and hard-to-soft {\rm \&} soft-to-hard diffusion.}
   \label{Huygens}
\end{figure}

The above discussion of the contributions to the total cross section
from the DGLAP and non-DGLAP parts of the phase space can conveniently
be cast in the form of the Huygens principle. To the standard DGLAP 
leading log$Q^{2}$ (LL$Q^{2}$) approximation one only considers
the contribution from the restricted part of the
available transverse phase space (\ref{eq:3.1.5}).
The familiar Huygens principle for the homogeneous DGLAP LL$Q^{2}$ evaluation 
of parton densities in the $x_{bj}$-$Q^{2}$ plane is illustrated in 
fig.~\ref{Huygens}a: one starts with the boundary condition 
$p(x,Q_{0}^{2})$ as a function of $x$ at fixed $Q_{0}^{2}$, 
the evolution paths $(z,\tilde{Q}^{2})$ for the calculation 
of $p(x,Q^{2})$ shown in fig.~5a are confined to a rectangle $x \leq z 
\leq 1,~ Q_{0}^{2} \leq \tilde{Q}^{2}\leq Q^{2}$, the evolution is unidirectional
in the sense that there is no feedback on the $x$-dependence of $p(x,Q_{1}^{2})$ 
from the $x$-dependence of $p(x,Q_{2}^{2})$ at $Q_{2}^{2} \geq Q_{1}^{2}$.
In fig.~5a we show some examples of evolution paths which are 
kinematically allowed but neglected in the DGLAP approximation. 
Starting with about flat or slowly rising $G(x,Q_{0}^{2})$ one finds that
the larger $Q^{2}$, the steeper the small-$x$ rise of $G(x,Q^{2})$.

At $x\ll 1$ the DGLAP contribution from the restricted transverse 
phase space (\ref{eq:3.1.5}) no longer dominates 
the multiparton production cross sections, the restriction 
(\ref{eq:3.1.5}) must be lifted and the contribution to the
cross section from small $\bkappa_{i}^2$ and large $\bkappa_{i}^2\gsim Q^2$
can no longer be neglected. The Huygens principle for the 
homogeneous BFKL evolution is illustrated in fig.~\ref{Huygens}b: one starts with 
the boundary condition ${\cal F}(x_{0},Q^{2})$ as a function of $Q^{2}$
at fixed $x_{0}\ll 1$, the evolution paths $(z,\tilde{Q}^{2})$ 
for the calculation 
of $p(x,Q^{2})$ are confined to a stripe $x \leq z \leq x_{0}$, in contrast
the the unidirectional DGLAP evolution one can say that under BFKL evolution
the small-$x$ behaviour of $p(x,Q^{2})$ at large $Q^{2}$ is fed partly by 
the $x$-dependence of soft  $p(x,Q^{2})$ at larger $x$ and vice versa. 
The most dramatic consequence of this soft-to-hard and 
hard-to-soft diffusion which can
not be eliminated is that  at very small $x$ the $x$-dependence 
of the gluon structure
in the soft and hard regions will eventually be the same:
\be
\lim_{{1\over x} \to \infty} G(x,Q^{2}) = G(Q^{2})
\left({1\over x}\right)^{\Delta_{\Pom}}\,.
\label{eq:3.3.2}
\ee
The rate of such a hard-to-soft diffusion is evidently sensitive to the infrared 
regularization of pQCD, the model estimates show that in the HERA range
of $x$ it is very slow \cite{NZZ94,NZBFKL}.


\section{The Ansatz for differential gluon structure function}

The major insight into parameterization of DGSF comes from early experience
with color dipole phenomenology of small-$x$ DIS. In color dipole approach,
which is closely related to $\bkappa$-factorization, the principal
quantity is the total cross section of interaction of the $q\bar{q}$ 
color dipole $\br$ with the proton target \cite{NZ94,NZglue,BGNPZUnit}
\be
\sigma(x,r)=
\frac{\pi^{2} r^{2}}{3}
\int 
\frac{d \bkappa^{2}}{\bkappa^{2}}
\frac{4[1-J_{0}(\kappa r)]}{(\kappa r)^{2}}
\alpha_{S}\left({\rm max}\{\bkappa^{2}, {A\over r^2}\}\right)
{\cal F}(x,\bkappa^{2}) \,,
\label{eq:4.00}
\ee
which for very small color dipoles can be approximated by 
\be
\sigma(x,r)=
\frac{\pi^{2} r^{2}}{3}\alpha_{S}\left({A\over r^2}\right)
G\left(x,{A\over r^2}\right) \,,
\label{eq:4.0}
\ee
where $A\approx 10$ comes from properties of the Bessel function 
$J_{0}(z)$.  
The phenomenological properties of the dipole cross section 
are well understood, for extraction of $\sigma(x,r)$ from the 
experimental data see \cite{NNPZsigmadipole,NNPZVM}.
The known dipole size dependence of $\sigma(x,r)$ serves as a
constrain on the possible 
$\bkappa^2$-dependence of ${\cal F}(x,\bkappa^{2})$.

As we argued in section 3.2, DGLAP fits are likely to overestimate
${\cal F}_{hard}(x,\bkappa^2)$ at moderate $\bkappa^{2}$.
Still, approximation (\ref{eq:4.0}) does a good job when
the hardness $A/r^2$ is very large, and at
large $Q^{2}$ we can arguably approximate the DGSF by the direct
differentiation of available fits (GRV, CTEQ, MRS, ...) to the integrated
gluon structure function $G_{pt}(x,Q^{2})$:
\be
{\cal F}_{pt}(x,\bkappa^{2}) \approx {\partial G_{pt}(x,\bkappa^{2}) \over 
\partial \log \bkappa^{2}}
\label{eq:4.1}
\ee
Hereafter the subscript $pt$  serves as a reminder that these gluon distributions
were obtained from the pQCD evolution analyses of the proton structure
function and cross sections of related hard processes. 

The available DGLAP fits are only applicable at $\bkappa^{2}\geq Q_{c}^{2}$,
see table 1 for the values of $Q_c^{2}$, in the extrapolation to
soft region $\bkappa^{2}\leq Q_{c}^{2}$ we are bound to educated guess.
 To this end recall that perturbative
gluons are confined and do not propagate to large distances; 
recent fits \cite{LatticeQCD} to the lattice QCD data suggest 
Yukawa-Debye screening of perturbative color fields with propagation/screening
radius $R_{c}\approx 0.27 fm$. 
Incidentally, precisely this value of $R_{c}$ for Yukawa screened 
colour fields has been used since 1994 in the very successful color 
dipole phenomenology of small-$x$  DIS \cite{NZHERA,BFKLRegge}. 
Furthermore, important finding of \cite{BFKLRegge} 
is a good quantitative description 
of the rising component of the proton structure function  
starting with the Yukawa-screened perturbative two-gluon exchange as a boundary
condition for the color dipole BFKL evolution.

The above suggests that $\bkappa^{2}$ dependence of perturbative hard
${\cal F}_{hard}(x,\bkappa^{2})$ in the soft region $\bkappa^{2}\leq Q_{c}^{2}$ 
is similar to the Yukawa-screened flux of photons in the 
positron, cf. eq.~(\ref{eq:2.3}),
with $\alpha_{em}$ replaced by the running
strong coupling of quarks $C_{F}\alpha_{S}(\bkappa^{2})$ and
with factor $N_c$ instead of 2 leptons in the positronium,
for the early discussion see \cite{NZsplit},
\be
{\cal F}^{(B)}_{pt}(\bkappa^2) =
C_F N_c {\alpha_{s}(\bkappa^2) \over \pi} \left( {\bkappa^2 \over 
\bkappa^2 +\mu_{pt}^{2}}\right)^{2} V_{N}(\bkappa)\,,
\label{eq:4.2}
\ee
Here $\mu_{pt}={1\over R_{c}}=0.75$ GeV  is the inverse Yukawa 
screening radius and must not be interpreted as a gluon mass; 
more sophisticated forms of screening can well be considered. 
Following \cite{NZHERA,BFKLRegge,NZ91,NZZ94} 
we impose also the infrared freezing of strong coupling:
$\alpha_{S}(\bkappa^2)\leq 0.82$; recently the concept of freezing coupling
has become very popular, for the review see \cite{Freezing}.

The vertex function $V_{N}(\bkappa)$ 
describes the decoupling of soft gluons, $\bkappa \ll {1\over R_{p}}$, 
from color neutral proton and has the same structure as in 
eq.~(\ref{eq:2.4}). In the nonrelativistic oscillator model 
for the nucleon one can  
relate the two-quark form factor of the nucleon to the
single-quark form factor,
\be
F_{2}(\vec{\bkappa},-\vec{\bkappa})=F_{1}\left({2N_{c} \over N_{c}-1}
\bkappa^{2}\right)\, .
\label{eq:4.3}
\ee
To the extent that $R_{c}^{2} \ll R_{p}^{2}$ the detailed functional 
form of $F_{2}(\vec{\bkappa},-\vec{\bkappa})$ is not crucial, the simple
relation (\ref{eq:4.3}) will be used also for a more realistic 
dipole approximation 
\be
F_{1}(\bkappa^{2})={1\over (1 + {\bkappa^2 \over \Lambda^{2}})^{2}}\,.
\label{eq:4.4}
\ee
The gluon probed radius of the proton and the charge radius of the proton
can be somewhat different and $\Lambda \sim 1$ GeV must be regarded as
a free parameter. 
Anticipating the forthcoming discussion of the
diffraction slope in vector meson production we put $\Lambda=1$ GeV.

As discussed above, the hard-to-soft diffusion makes the DGSF rising at
small-$x$ even in the soft region. We model this hard-to-soft diffusion 
by matching the $\bkappa^{2}$ dependence (\ref{eq:4.2}) to the DGLAP 
fit $ {\cal F}_{pt}(x,Q_{c}^{2})$ at the soft-hard interface $Q_{c}^{2}$
and assigning to ${\cal F}_{hard}(x,\bkappa^2)$ in the region of 
$\bkappa^{2}\leq Q_{c}^{2}$ the same $x$-dependence 
as shown by the DGLAP fit $ {\cal F}_{pt}(x,Q_{c}^{2})$, i.e.,
\be
{\cal F}_{hard}(x,\bkappa^2)= 
{\cal F}^{(B)}_{pt}(\bkappa^2){{\cal F}_{pt}(x,Q_{c}^{2})
\over {\cal F}_{pt}^{(B)}(Q_{c}^{2})}
\theta(Q_{c}^{2}-\bkappa^{2}) +{\cal F}_{pt}(x,\bkappa^2)
\theta(\bkappa^{2}-Q_{c}^{2})\,.
\label{eq:4.5}
\ee
Because the accepted propagation radius $R_{c} \sim 0.3$ fm for perturbative 
gluons is short compared to a typical range of strong
interaction, the dipole cross section (\ref{eq:4.00}) evaluated with
the DGSF (\ref{eq:4.5}) would miss an interaction strength in the soft region,
for large color dipoles.

In Ref.\cite{NZHERA,BFKLRegge} interaction of large dipoles has been modeled 
by the non-perturbative, soft mechanism with energy-independent dipole cross 
section, whose specific form \cite{NZHERA,JETPVM} has been driven by early
analysis \cite{NZ91} of the exchange by two nonperturbative gluons.
More recently several closely related models for $\sigma_{\rm soft}(r)$ have
appeared in the literature, see for instance models for dipole-dipole
scattering via polarization of non-perturbative QCD vacuum \cite{Nachtmann}
and the model of soft-hard two-component pomeron \cite{LANDSH}. In order
to reproduce the required interaction strength for large dipoles, 
we introduce the genuinely soft, nonperturbative
component of DGSF which we parameterize as
\be
{\cal F}^{(B)}_{soft}(x,\bkappa^2) = a_{soft}
C_F N_c {\alpha_{s}(\bkappa^2) \over \pi} \left( {\bkappa^2 \over 
\bkappa^2 +\mu_{soft}^{2}}\right)^2 V_{N}(\bkappa)\,.
\label{eq:4.6}
\ee
The principal point about this non-perturbative component of DGSF is that 
it must not be subjected to pQCD evolution. Thus the arguments about the 
hard-to-soft diffusion driven rise of perturbative DGSF even 
at small $\bkappa^{2}$ do not apply to the non-perturbative DGSF 
and we take it energy-independent one. Such a nonperturbative component 
of DGSF would have certain high-$\bkappa^{2}$ tail which should not
extend too far. The desired suppression of soft DGSF at large-$\bkappa^{2}$
and of hard DGSF (\ref{eq:4.5}) at moderate and small $\bkappa^{2}$ 
can be achieved by the extrapolation of the form suggested 
in \cite{NPZLT,Twist4}
\be
{\cal F}(x,\bkappa^2)= {\cal F}^{(B)}_{soft}(x,\bkappa^2) 
{\bkappa_{s}^2 \over 
\bkappa^2 +\bkappa_{s}^2} + {\cal F}_{hard}(x,\bkappa^2) 
{\bkappa^2 \over 
\bkappa^2 +\bkappa_{h}^2} 
\label{eq:4.7}
\ee
The above described Ansatz for DGSF must be regarded as a poor man's approximation.
The separation of small-$\bkappa^{2}$ DGSF into the genuine nonperturbative
component and small-$\bkappa^{2}$ tail of the hard perturbative DGSF 
is not unique. Specifically, we attributed to the latter the same small-$x$
rise as in the DGLAP fits at $Q_{c}^{2}$ while one can not exclude that the 
hard DGSF has a small $x$-independent component. The issue of soft-hard
separation must be addressed in dynamical models for infrared regularization
of perturbative QCD. As we shall see below, in section 4.4, the soft component 
of the above described Ansatz is about twice larger than the soft component 
used in the early color dipole phenomenology \cite{NZHERA,BFKLRegge}.

The $\bkappa$-factorization formulas (\ref{eq:3.1.7}) and (\ref{eq:3.1.8})
correspond to the full-phase space extension of the LO DGLAP approach at
small $x$. For this reason our Ans\"atze for ${\cal F}_{hard}(x,Q^{2})$
will be based on LO DGLAP fits to the gluon structure function of the
proton $G_{pt}(x,Q^2)$. We consider the GRV98LO \cite{GRV}, CTEQ4L, 
version 4.6 \cite{CTEQ} and MRS LO 1998 \cite{MRS} parameterizations. 
We take the liberty of referring to our Anz\"atze for DGSF based on those 
LO DGLAP input as D-GRV, D-CTEQ and D-MRS parameterizations, respectively. 

Our formulas (\ref{eq:3.1.5}), (\ref{eq:3.1.6}) describe the sea component 
of the proton structure function. Arguably these LL${1\over x}$ 
formulas are applicable at $x\lsim x_{0}= 1\div 3 \cdot 10^{-2} $.
At large $Q^{2}$ the experimentally attainable values of $x$ are not so
small. In order to give a crude idea on finite-energy effects at moderately
small $x$, we stretch our fits to $x\gsim x_{0}$ multiplying the above 
Ansatz for DGSF by the purely phenomenological factor $(1-x)^{5}$ motivated 
by the familiar large-$x$ behaviour of DGLAP parameterizations of the gluon 
structure function of the proton. We also add to the sea components 
(\ref{eq:3.1.5}), (\ref{eq:3.1.6}) the contribution from DIS on valence quarks
borrowing the parameterizations from the respective GRV, CTEQ and MRS fits.
The latter are only available for $Q^2 \geq Q_{c}^2$.
At $x\lsim 10^{-2}$ this valence contribution is small and fades away rapidly
with decreasing $x$, for instance see \cite{BFKLRegge}.


\section{Determination of the differential gluon structure function of the proton}


\subsection{The parameters of DGFS for different DGLAP inputs}

Our goal is a determination of small-$x$ DGSF in the whole range of 
$\bkappa^{2}$ by adjusting the relevant parameters to the experimental 
data on small-$x$ $F_{2p}(x,Q^2)$ in the whole available region of $Q^2$ 
as well as the real photoabsorption cross section. The theoretical
calculation of these observables is based on Eqs.~(\ref{eq:3.1.7}),
~(\ref{eq:3.1.8}),~(\ref{eq:4.7}).

The parameters which we did not try adjusting but
borrowed from early work in the color dipole picture are $R_{c}=0.27$ fm, 
i.e., $\mu_{pt}=0.75$ GeV and the frozen value of the LO QCD coupling
with $\Lambda_{QCD}=0.2$ GeV:
\be
\alpha_{S}(Q^{2})=
{\rm min}\left\{0.82,{4\pi\over \beta_{0}\log{Q^{2}\over 
\Lambda_{QCD}^{2}}}\right\}\, .
\label{eq:5.1}
\ee
We recall that the GRV, MRS and CTEQ fits
to GSF start the DGLAP evolution at quite a different soft-to-hard interface 
$Q_{c}^{2}$ and diverge quite a lot, especially at moderate and small 
$\bkappa^{2}$. The value of $Q_{c}^{2}$ is borrowed from these fits and
is not a free parameter.

The adjustable parameters are $\mu_{soft}$, $a_{soft}$, $m_{u,d}$, 
$\bkappa_s^2$ and $\bkappa_h^2$. We take $m_s=m_{u,d}+0.15 $GeV and
$m_{c}=1.5$ GeV. The r\^ole of these parameters is as follows.
The quark mass $m_{u,d}$ defines the transverse size of the $q\bar{q}
=u\bar{u},d\bar{d}$ Fock state of the photon, whereas $\mu_{soft}^{-2}$ 
controls the $r$-dependence of, and in conjunction with $a_{soft}$
sets the scale for, the dipole cross section for large 
size $q\bar{q}$ dipoles in the photon. The both $m_{u,d}$ 
and $\mu_{soft}$ have clear physical meaning and we have certain 
insight into their variation range form the early work on color dipole
phenomenology of DIS. The 
magnitude of the dipole cross section at large and moderately small 
dipole size depends also on the soft-to-hard interpolation of DGSF, 
which is somewhat different for different LO DGLAP inputs for 
$G_{pt}(x,Q^{2})$. This difference of DGLAP inputs can be corrected 
for by adjusting $\mu_{soft}^{2}$ and the value of $\bkappa_h^2$. 
Because of soft-to-hard and hard-to-soft diffusion the DGLAP fits
are expected to fail at small $x$, we allow for $x$ dependence of
$\bkappa_h^2$. 

Evidently, roughly equal values 
of $F_{2p}(x,Q^2)$ can be obtained for somewhat smaller ${\cal F}(x,Q^{2})$ 
at the expense of taking smaller $m_{u,d}$ and vise versa. 
Therefore, though the quark mass does not explicitly 
enter the parameterization  for ${\cal F}(x,\bkappa^2)$,
the preferred value of $m_{u,d}$ turns out to be correlated with the 
Ansatz for DGSF, i.e. each particular parameterization
of DGSF implies certain $m_{u,d}$.
In what follows we set $a_{soft} = 2$, $\bkappa_s^2 = 
3.0$ GeV$^2$, $m_{u,d} = 0.22$ GeV, 
so that only $\bkappa_h^2$ and $\mu_{soft}$ varied from one DGLAP input 
to another, see Table 1. The soft components of the D-GRV and D-CTEQ
parameterizations turn out identical. 
The eye-ball fits are sufficient for the
purposes of the present exploratory study. The parameters found are 
similar to those used in \cite{NPZLT,Twist4} where the focus has been
on the description of diffractive DIS. 

\begin{center}
{Table 1. The parameters of differential gluon structure function
for different DGLAP inputs.\vspace{0.3cm}\\}

 \begin{tabular}{|r|c|c|c|}
\hline
& D-GRV & D-MRS & D-CTEQ  \\ \hline\hline
 LO DGLAP input & GRV98LO \cite{GRV}&  CTEQ4L(v.4.6) \cite{CTEQ} &
MRS-LO-1998 \cite{MRS}\\
$Q_c^2$, GeV$^2$            & 0.895 & 1.37 & 3.26  \\
$\kappa_h^2$, GeV$^2$ & $\left(1 + 0.0018\log^4{1\over x}\right)^{1/2}$ 
& $\left(1 + 0.038\log^2{1\over x}\right)^{1/2}$ & 
$\left(1 + 0.047\log^2{1\over x} \right)^{1/2}$  \\
$\mu_{soft}$, GeV     & 0.1 & 0.07 & 0.1\\ \hline
 \end{tabular}
\end{center}

One minor problem encountered in numerical differentiation of all 
three parameterizations for $G_{pt}(x,Q^2)$ 
was the seesaw $\bkappa^2$-behavior of the resulting DGSF (\ref{eq:4.1}),
which was an artifact of the grid interpolation routines. Although 
this seesaw behavior of DGSF is smoothed out in integral observables
like $G(x,Q^2)$ or $F_{2p}(x,Q^2)$, we still preferred to remove 
the unphysical seesaw cusps and have smooth DGSF.
This was achieved by calculation DGSF from (\ref{eq:4.1})
at the center of each interval of the $Q^2$-grid and further interpolation
of the results between these points. By integration of the so-smoothed
${\cal F}_{pt}(x,Q^2)$ one recovers the input $G_{pt}(x,Q^2)$.
The values of $Q_c^2$ cited in Table 1 corresponds to centers of the
first bin of the corresponding $Q^{2}$-grid.

\begin{figure}[!htb]
   \centering
   \epsfig{file=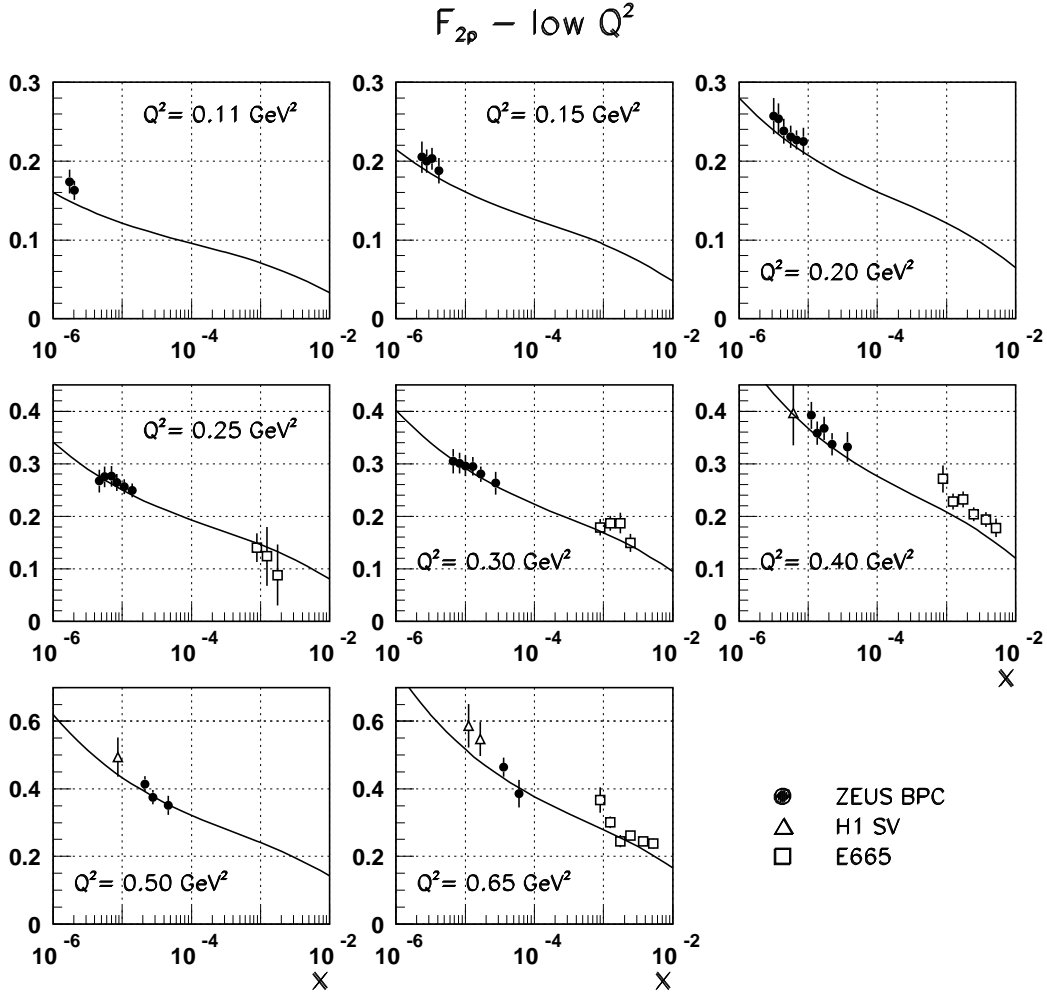,width=16cm}
   \caption{\em The $\bkappa$-factorization description
of the experimental data on $F_{2p}(x,Q^2)$ in the low $Q^2$ region;
black circles are ZEUS BPC data \cite{ZEUSBPC}, 
open triangles denote H1 shifted vertex (SV) data \cite{H1shifted},
open squares are E665 data \cite{E665}.
Solid line represents $\bkappa$-factorization results for the D-GRV parameterization
of the differential gluon structure function ${\cal F}(x,\bkappa^2)$.}
   \label{F2protonLowQ2}
\end{figure}


\subsection{The description of the proton structure function $F_{2p}(x,Q^2)$ }

We focus on the sea dominated leading log${1\over x}$ region of $x< 10^{-2}$. 
The practical calculation of the proton structure function involves the 
two running arguments of DGSF: $x_{g}$ and $\bkappa^{2}$. We recall that in 
the standard collinear DGLAP approximation one has $\bkappa^{2} \ll k^{2} 
\ll Q^{2}$ and $x_{g}\approx 2x$, see eq.~(\ref{eq:3.2.1}). 
Within the $\bkappa$-factorization one finds that the
dominant contribution to $F_{2p}(x,Q^{2})$ comes from $M_{t}^{2}\sim Q^{2}$
with little contribution from $M_{t}^{2}\gsim Q^{2}$.
Because at small $x_g$ the $x_g$ dependence of ${\cal F}(x_g,Q^{2})$ is 
rather steep, we take into account the $x_{g}-x_{bj}$ relationship 
(\ref{eq:3.1.11}). Anticipating the results on effective intercepts 
to be reported in section 7, we 
notice that for all practical purposes one can neglect the impact of $\bkappa$
on the relationship (\ref{eq:3.1.11}), which simplifies greatly the numerical
analysis. Indeed, the $x_{g}$ dependence of
${\cal F}(x_g,\bkappa^{2})$ is important only at large $\bkappa^{2}$, which
contribute to $F_{2p}(x,Q^{2})$ only at large $Q^{2}$; but the larger 
$Q^{2}$, the better holds the DGLAP ordering $\bkappa^{2} \ll k^{2}, Q^{2}$. 
Although at small to moderate $Q^{2}$ the DGLAP the ordering breaks down, 
the $x_{g}$ dependence of ${\cal F}(x_g,\bkappa^{2})$ is negligible
weak here.

\begin{figure}[!htb]
   \centering
   \epsfig{file=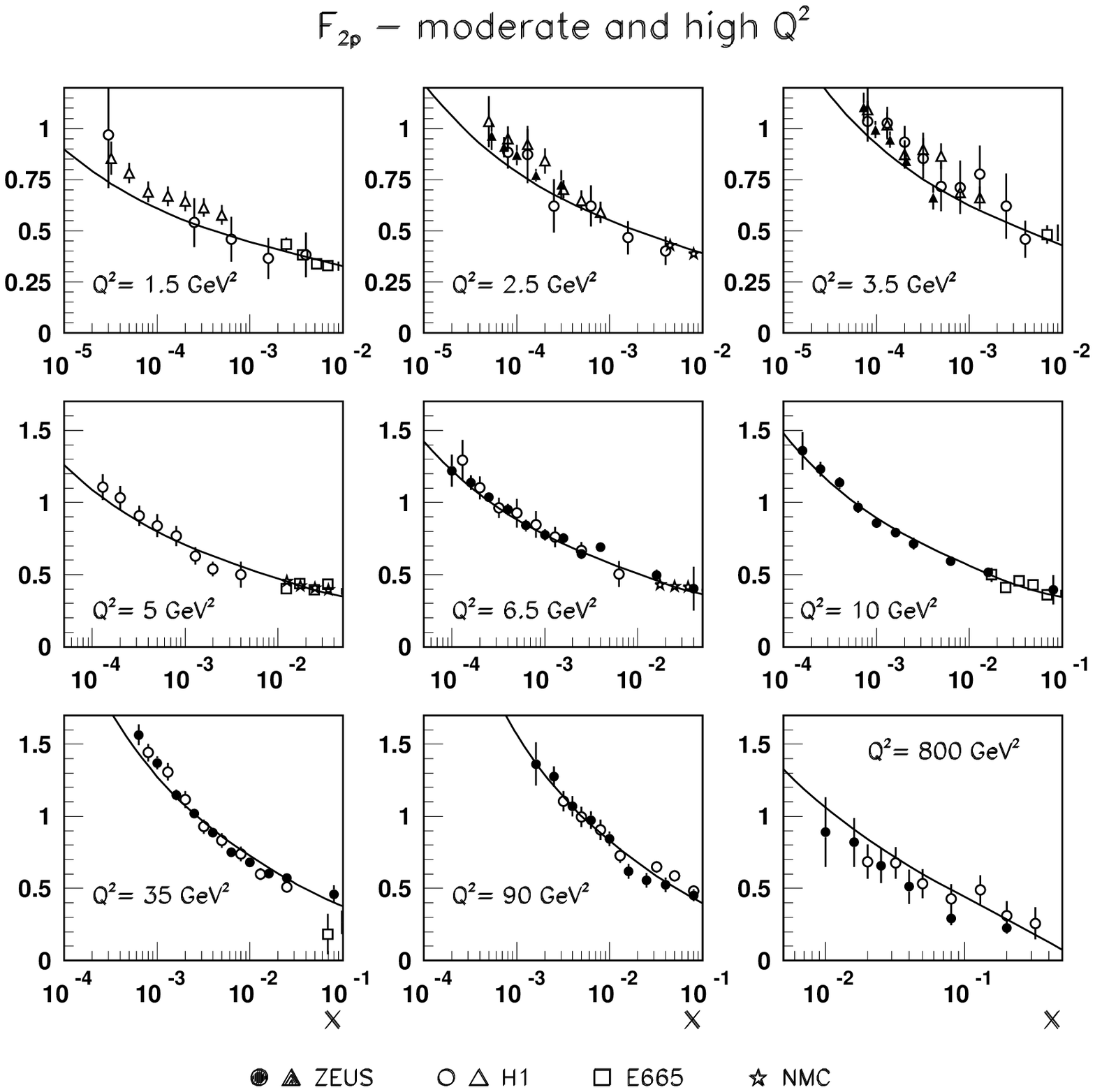,width=16cm}
   \caption{\em The $\bkappa$-factorization description
of the experimental data on $F_{2p}(x,Q^2)$ in the 
moderate and high $Q^2$ region;
black circles and triangles are ZEUS data \cite{ZEUSlarge}, 
\cite{ZEUSshifted}, open circles and triangles 
show H1 data \cite{H1large}, \cite{H1shifted}, 
open squares are E665 data \cite{E665},
stars refer to NMC results \cite{NMC}.
Solid line represents $\bkappa$-factorization results for the D-GRV parameterization
of the differential gluon structure function ${\cal F}(x,\bkappa^2)$.}
   
\label{F2protonLargeQ2}
\end{figure}

\begin{figure}[!htb]
   \centering
   \epsfig{file=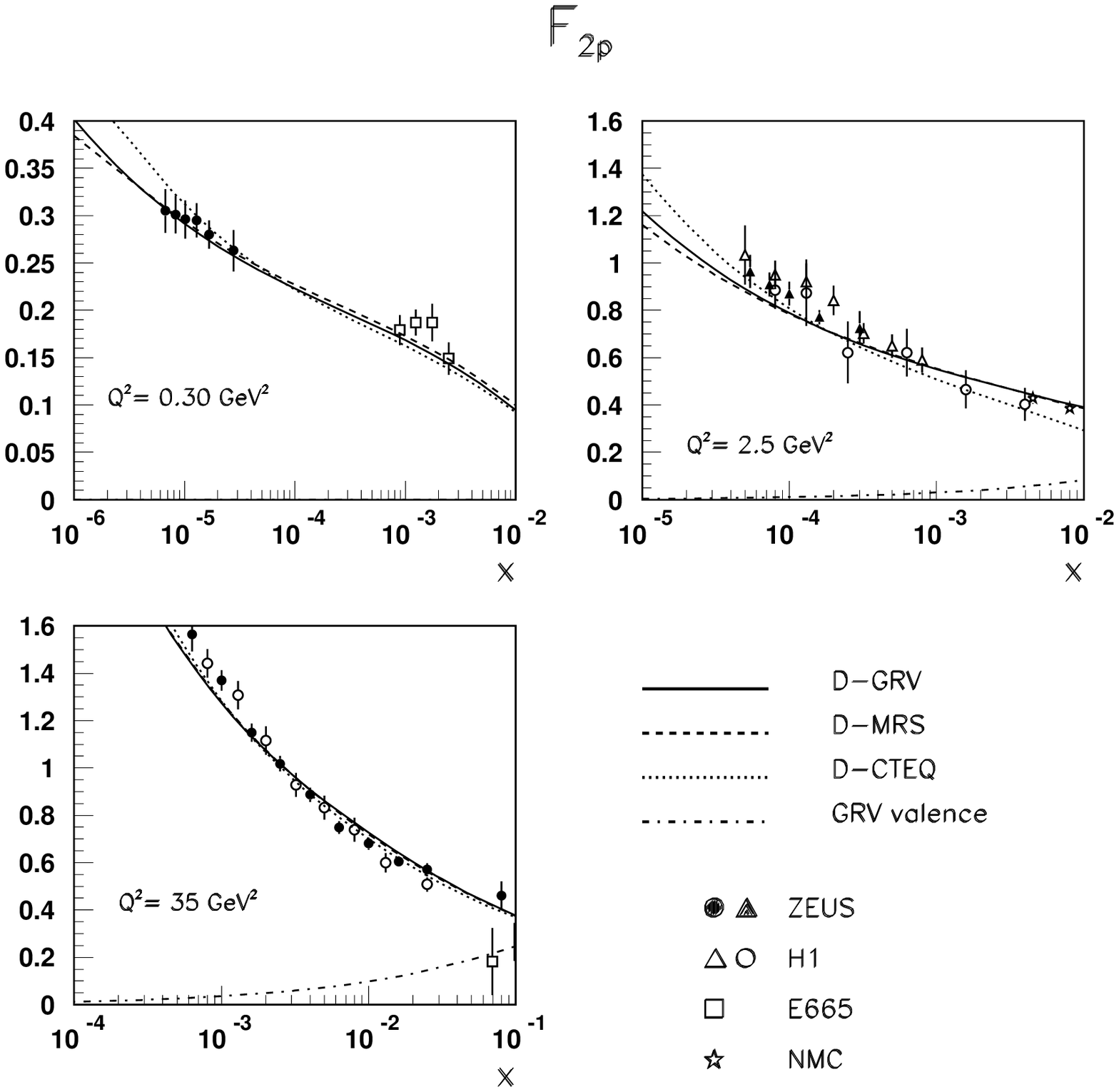,width=14cm}
   \caption{\em A comparison of the $\bkappa$-factorization description
of the experimental data on $F_{2p}(x,Q^2)$ for several values of
$Q^{2}$ with the D-GRV, D-CTEQ and D-MRS 
parameterizations of the differential gluon structure 
function ${\cal F}(x,\bkappa^2)$. The contribution to $F_{2p}(x,Q^2)$
from DIS off valence quarks is shown separately for larger $Q^2$.}
   \label{F2pcompare}
\end{figure}

As we shall discuss in more detail below, achieving a good agreement
from small to moderate to large $Q^{2}$ is a highly nontrivial task,
because strong modification of the soft contribution to ${\cal F}(x,Q^2)$ 
unavoidably echos in the integral quantity $G_{D}(x, Q^2)$
throughout the whole range of $Q^2$ and shall affect the
calculated structure function from small to moderate to large $Q^{2}$.

The quality of achieved description of the experimental data
on the small-$x$ proton structure function is illustrated by
figs.~\ref{F2protonLowQ2},~\ref{F2protonLargeQ2}. 
The data shown include recent HERA data
(ZEUS \cite{ZEUSlarge}, ZEUS shifted vertex \cite{ZEUSshifted}, 
ZEUS BPC \cite{ZEUSBPC}, H1 \cite{H1large}, 
H1 shifted vertex \cite{H1shifted}),
FNAL E665 experiment \cite{E665} and CERN NMC experiment \cite{NMC}.
When plotting the E665 and NMC data, we took the liberty of shifting
the data points from the reported
values of $Q^2$ to the closest $Q^2$ boxes for which
the HERA data are available. For $Q^{2} < Q_{c}^2 =0.9$ GeV$^2$
the parameterizations for valence distributions are not available
and our curves show only the sea component of $F_{2p}(x,Q^2)$,
at larger $Q^{2}$ the valence component is included.

\begin{figure}[!htb]
   \centering
   \epsfig{file=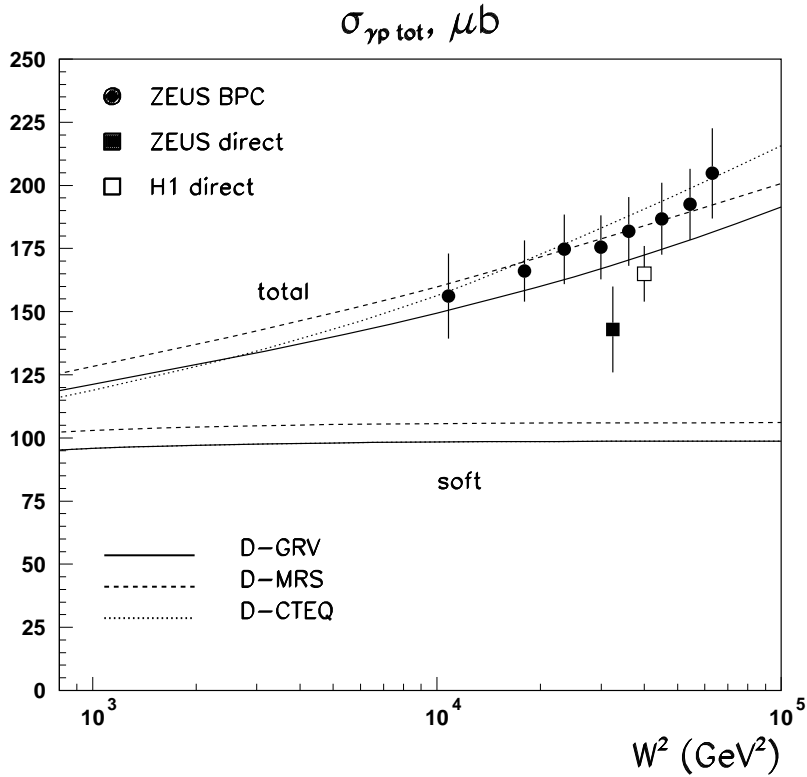,width=12cm}
   \caption{\em A comparison of the $\bkappa$-factorization description
of the experimental data on real photoabsorption cross section
based on the D-GRV, D-CTEQ and D-MRS 
parameterizations  of the differential gluon structure 
function ${\cal F}(x,\bkappa^2)$. The squares
show the experimental data from 1992-93 direct measurements, the
bullets are the results of extrapolation of virtual 
photoabsorption to $Q^{2}=0$ (\cite{ZEUSBPC} and references therein). 
The soft component of 
photoabsorption cross section is shown separately. }
   \label{RealPhoton}
\end{figure}

At $x< 10^{-2}$ the accuracy of our D-GRV description
of the proton structure function is commensurate to 
that of the accuracy of standard LO GRV fits. In order not to
cram the figures with nearly overlapping curves, we show the results 
for D-GRV parameterization. The situation with
D-CTEQ and D-MRS is very similar, which is seen in fig.~\ref{F2pcompare}, 
where we show on a larger scale simultaneously the results 
from the D-GRV, D-CTEQ and D-MRS DGSFs for several selected values 
of $Q^{2}$. Here at large $Q^2$ we show separately the contribution from
valence quarks. 
The difference between the results for $F_{2p}(x,Q^2)$ 
for different DGLAP inputs is marginal for all the practical purposes,
see also a comparison of the results for $\sigma^{\gamma p}$  
for different DGLAP inputs in fig.~\ref{RealPhoton}.


\subsection{Real photoabsorption cross section $\sigma^{\gamma p}$.}

In the limiting case of $Q^2=0$ the relevant observable is the real 
photoabsorption cross section $\sigma^{\gamma p}$. Although the
Bjorken variable is meaningless at very small $Q^{2}$, the
gluon variable $x_{g}$ remains well defined at $Q^{2}=0$, 
see eq.~(\ref{eq:3.1.11}). In fig.~\ref{RealPhoton} we present 
our results alongside with the results of the direct measurements 
of $\sigma^{\gamma p}$ and the results of extrapolation of virtual
photoabsorption cross sections to $Q^{2}=0$,
for the summary of the experimental
data see \cite{ZEUSBPC}. We emphasize that
we reproduce well the observed magnitude and pattern of the energy dependence  
of $\sigma^{\gamma p}$ in an approach with the manifestly energy-independent
soft contribution to the total cross section (which is shown separately 
in fig.~\ref{RealPhoton}). We recall that our parameterizations for 
${\cal F}(x,Q^2)$ give identical soft cross sections for the
GRV and CTEQ inputs (see table 1). 
The barely visible decrease of $\sigma^{\gamma p}_{soft}$
towards small $W$ is a manifestation of $\propto (1-x)^5$ large-$x$
behaviour of gluon densities. The extension to lower energies requires
introduction of the secondary reggeon exchanges which goes beyond
the subject of this study.

In our scenario the energy dependence of
$\sigma^{\gamma p}$ is entirely due to the $x_g$-dependent hard component 
${\cal F}_{hard}(x_g,Q^2)$ and as such this rise of the total cross
section for soft reaction can be regarded as driven entirely by 
very substantial hard-to-soft diffusion. Such a scenario has repeatedly
been discussed earlier \cite{NZHERA,BFKLRegge,VMscan}. Time and time again
we shall see similar effects of hard-to-soft diffusion and vise versa.  
Notice that hard-to-soft diffusion 
is a straightforward consequence of full phase space calculation of
partonic cross sections and we do not see any possibility for decoupling of
hard gluon contribution from the total cross sections of 
any soft interaction, whose generic example is the real photoabsorption.


\section{Properties of differential gluon structure
function in the momentum space}


\subsection{Soft/hard decomposition of DGSF}

\begin{figure}[!htb]
   \centering
   \epsfig{file=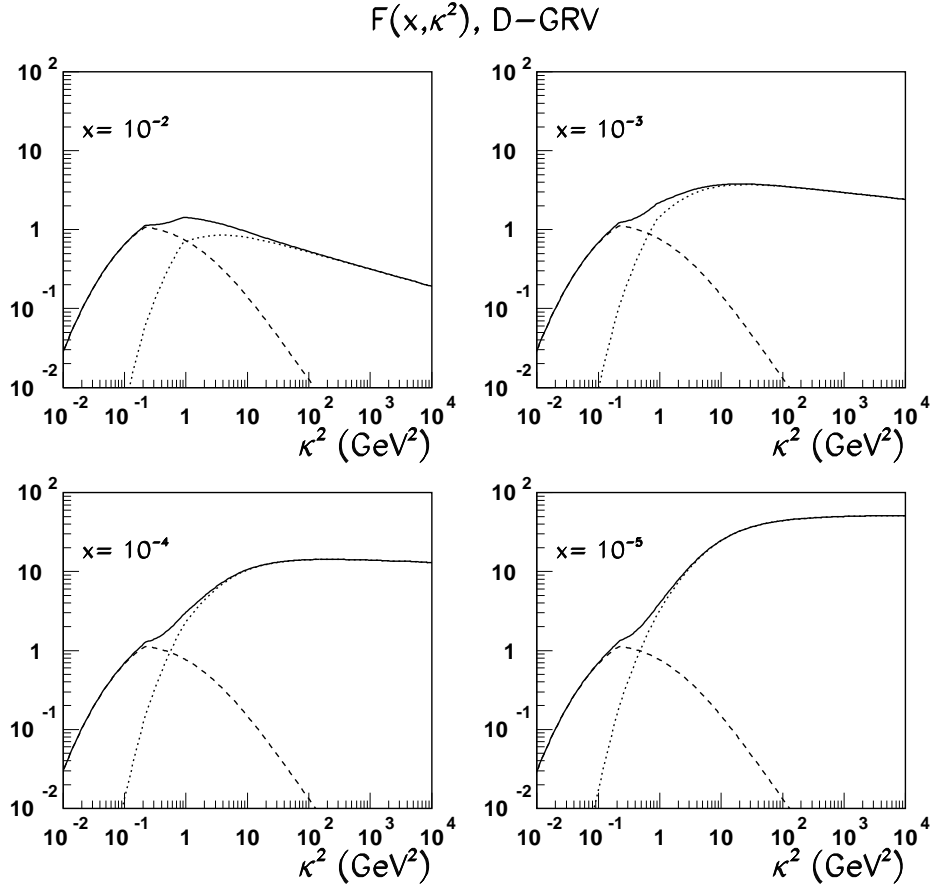,width=14cm}
   \caption{\em D-GRV differential gluon structure function
${\cal F}(x,\bkappa^2)$ as a function of 
$\bkappa^2$ at several values of $x$. Dashed and dotted 
lines represent the soft and hard components; the total 
unintegrated gluon density is shown by the solid line}
   \label{DGSF}
\end{figure}

\begin{figure}[!htb]
   \centering
   \epsfig{file=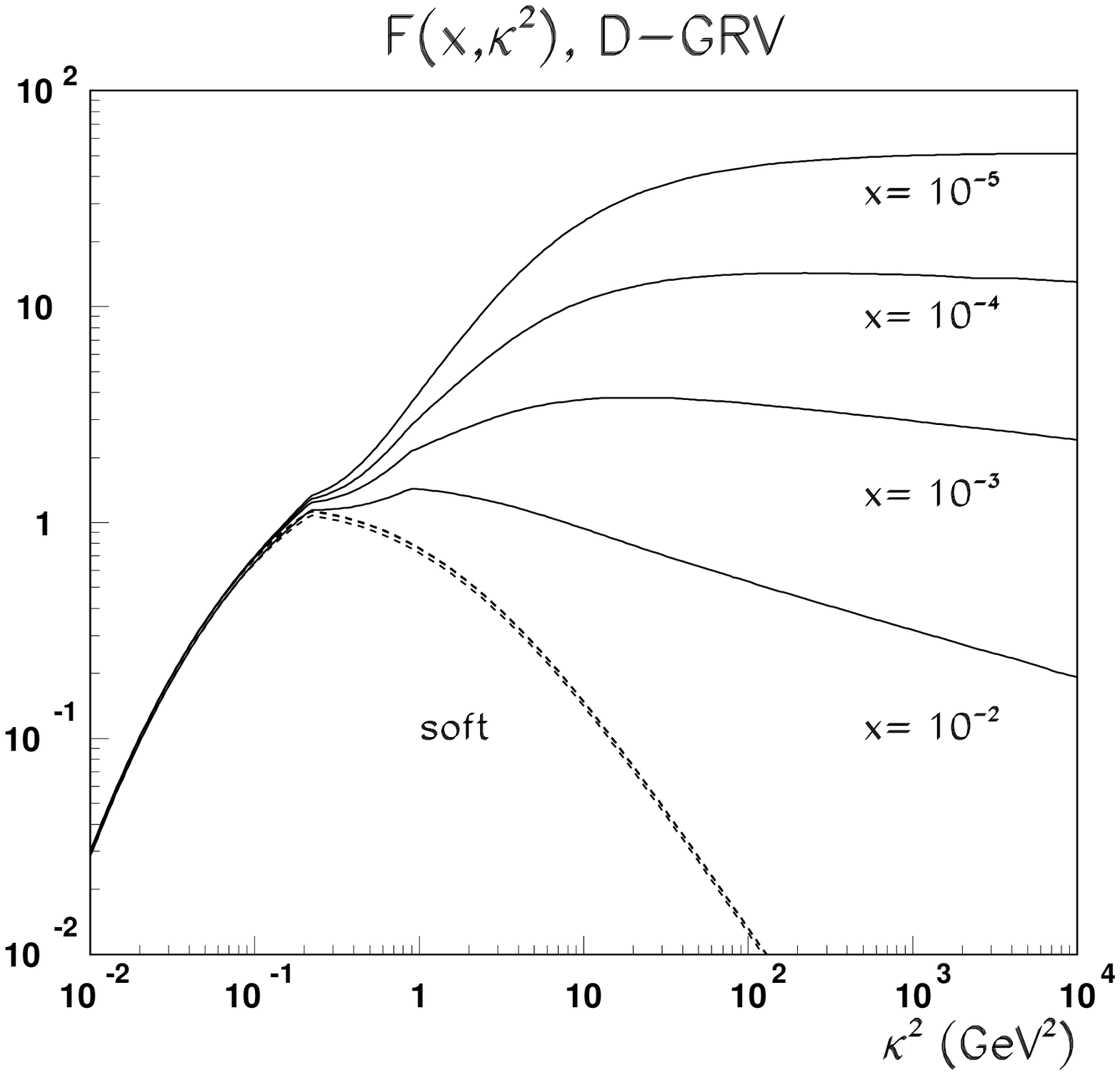,width=10cm}
   \caption{\em The same as in Fig.~\ref{DGSF} but overlaid onto one graph 
for illustration of the $x$-dependence of ${\cal F}(x,\bkappa^2)$. The dashed 
lines shows the soft component ${\cal F}_{soft}(x,\bkappa^2)$ and its slight
variation with $x$ due to the finite-$x$ factor $(1-x)^5$. }
   \label{DGSFoverlaid}
\end{figure}

\begin{figure}[!htb]
   \centering
   \epsfig{file=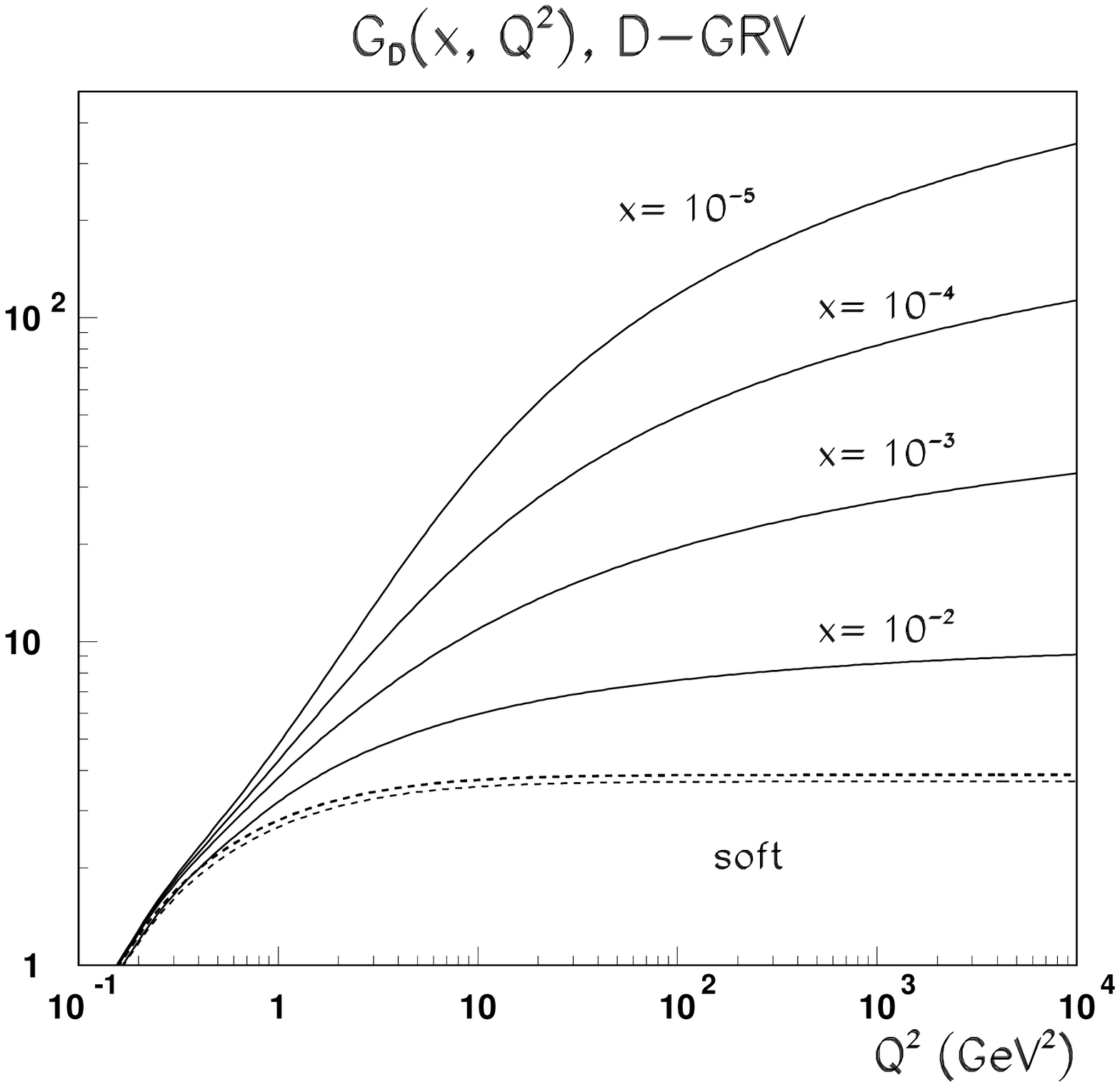,width=10cm}
   \caption{\em The same as Fig.~11 but for integrated gluon 
structure function $G_{D}(x,Q^2)$ as given by the D-GRV 
parameterization of the differential gluon structure 
function ${\cal F}(x,\bkappa^2)$, for the discussion see Section 6.2.}
   \label{GSF}
\end{figure}

Now we focus on the $x$ and $\bkappa^2$ behavior of the
so-determined DGSF starting for the reference with the D-GRV 
parameterization. The same pattern holds for DGSF 
based on CTEQ and MRS DGLAP inputs, see below.
In figs.~\ref{DGSF} and \ref{DGSFoverlaid} 
we plot the differential gluon density ${\cal F}(x_g,Q^2)$, 
while in fig.~\ref{GSF} we show the integrated gluon density 
\be
G_{D}(x,Q^{2})=\int^{Q^{2}}_{0} {d\kappa^{2} \over \kappa^{2}}
{\cal F}(x,\bkappa^{2})\, .
\label{eq:5.2.1}
\ee
Here the subscript {\sl D} is a reminder that the integrated
$G_{D}(x,Q^{2})$ is derived from DGSF. As such it
must not be confused with  the DGLAP parameterizations 
$G_{pt}(x,Q^{2})$ supplied with the subscript $pt$.

Figs.~\ref{DGSF} and \ref{DGSFoverlaid} illustrate 
the interplay of the nonperturbative soft 
component of DGSF and perturbative hard contribution supplemented 
with the above described continuation into $\bkappa^{2}\leq Q_{c}^2$
at various $x$ and $\bkappa^2$. The soft and hard contributions are 
shown by dashed and dotted lines respectively; 
their sum is given by solid line.

Apart from the large-$x$ suppression factor
$(1-x)^5$ our non-perturbative soft component does not depend on $x$. At a not so 
small $x=10^{-2}$ it dominates the soft region of 
$\bkappa^2 \lsim 1\div 2 $ GeV$^2$,
the hard component takes over at higher $\bkappa^2$. The  soft-hard
crossover point is close to $\mu_{pt}^{2}$ but because of the 
hard-to-soft diffusion it moves with decreasing $x$ 
to a gradually smaller $Q^{2}$. 

\subsection{Soft/hard decomposition of the integrated gluon structure function}

The r{\^o}le of the soft component if further illustrated by fig.~\ref{GSF}
where we show the integrated gluon density (\ref{eq:5.2.1})
and its soft and hard components $G_{soft}(x, Q^2)$ and $G_{hard}(x, Q^2)$,
respectively. The soft contribution $G_{soft}(x, Q^2)$ 
is a dominant feature of the integrated
gluon density $G_{D}(x, Q^2)$ for  $Q^{2}\lsim 1$ GeV$^2$.
It builds up rapidly with $Q^{2}$ and receives the major
contribution from the region $\bkappa^2 \sim 0.3 \div 0.5 $ GeV$^2$.
Our Ansatz for ${\cal F}_{soft}(x,\bkappa^2)$ is such that it starts decreasing
already at $\bkappa^2 \sim 0.2$ GeV$^2$ and  vanishes rapidly
beyond $\kappa^{2} \gsim \kappa_{soft}^2$, see figs. 10,11. Still the residual 
rise of the soft gluon density beyond $Q^2 \sim  0.5 $ GeV$^2$
is substantial: $G_{soft}(x, Q^2)$ rises by about the factor two 
before it flattens at large $Q^{2}$. We emphasize that 
$G_{soft}(Q^2)$ being finite at large $Q^{2}$ is quite natural
--- a decrease of $G_{soft}(Q^2)$ at large $Q^{2}$ only is possible
if ${\cal F}_{soft}(Q^{2})$ becomes negative valued at large $Q^2$, which
does not seem to be a viable option.

At moderately small $x=10^{-2}$ the scaling violations are still weak 
and the soft contribution $G_{soft}(x, Q^2)$ remains a substantial 
part, about one half, of integrated GSF $G_{D}(x, Q^2)$ {\em at all} $Q^{2}$.
At very small $x\lsim 10^{-3}$ the scaling violations in the gluon 
structure function are strong and $G_{hard}(x,Q^{2}) \gg G_{soft}(x,Q^{2})$ 
starting from $Q^{2} \sim$ 1-2 GeV$^2$.


\subsection{Soft/hard decomposition of the proton structure
function $F_{2}(x,Q^{2})$ }

\begin{figure}[!htb]
   \centering
   \epsfig{file=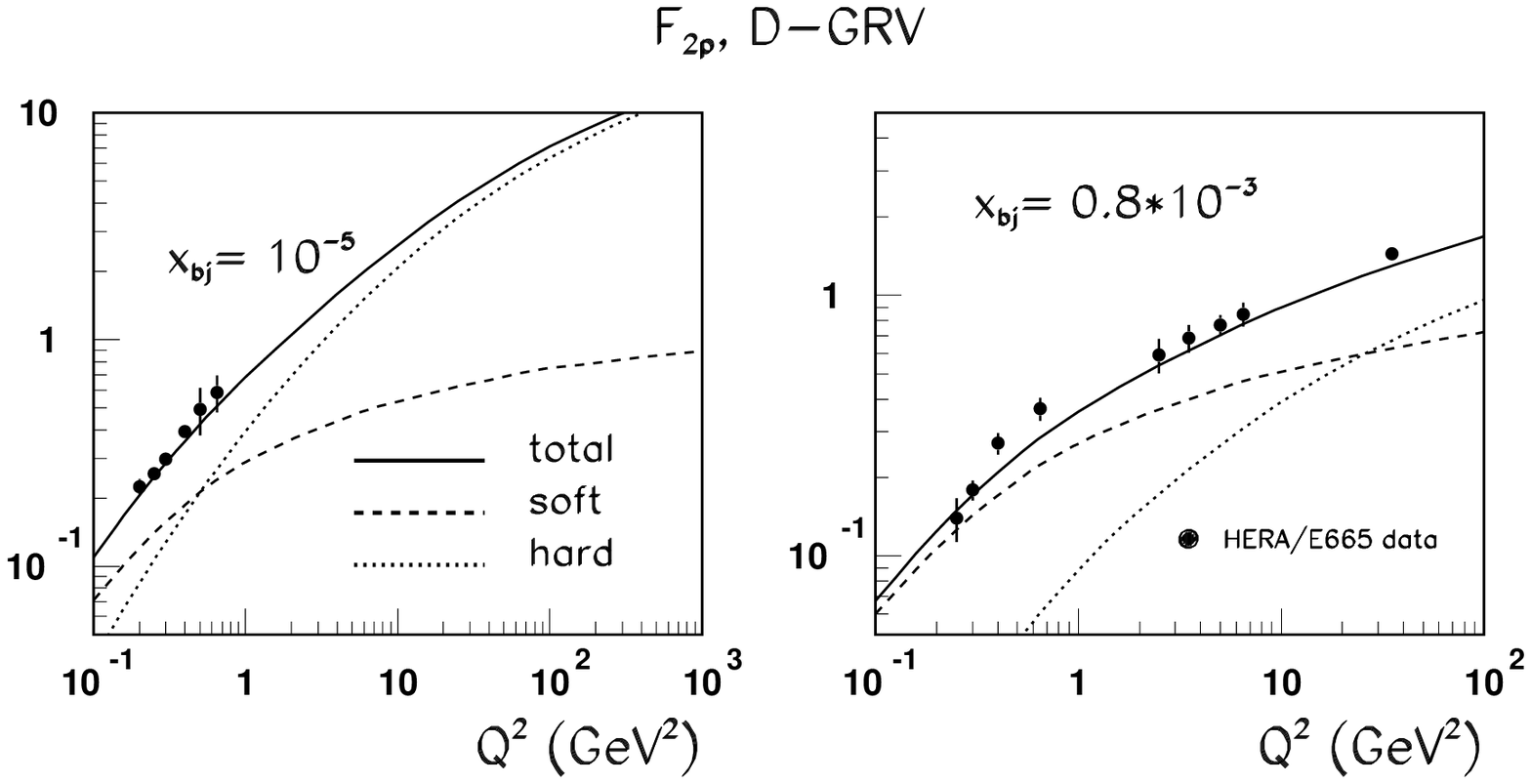,width=16cm}
   \caption{\em The soft-hard decomposition of $\bkappa$-factorization
results for the 
proton structure function $F_{2p}(x,Q^{2})$
evaluated with the
D-GRV parameterization of the differential gluon structure 
function ${\cal F}(x,\bkappa^2)$.}
   \label{Soft-Hard.F2p}
\end{figure}

Eqs. (\ref{eq:3.1.7}), (\ref{eq:3.1.8}) define 
the soft/hard decomposition of the proton structure 
function. In fig.~\ref{Soft-Hard.F2p} we show $F_{2p}^{hard}(x,Q^{2})$ and 
$F_{2p}^{soft}(x,Q^{2})$ as a function of $Q^{2}$ for the two
representative values 
of $x$. 
Notice how significance of soft component as a function
of $Q^{2}$ rises from fully
differential ${\cal F}(x,Q^{2})$ to integrated $G_{D}(x,Q^2)$ to
doubly integrated $F_{2p}^{soft}(x,Q^{2})$. At a moderately small
$x\sim 10^{-3}$ the soft contribution is a dominant part of
$ F_{2p}(x,Q^{2})$, although the rapidly rising hard component
$F_{2p}^{hard}(x,Q^{2})$ gradually takes over at smaller $x$.

Notice that not only does $F_{2p}^{soft}(x,Q^2)$ not vanish at large $Q^{2}$,
but also it rises slowly with $Q^{2}$ as
\be
F_{2p}^{soft}(x,Q^{2})\sim \sum e_{f}^2{4 G_{soft}(Q^{2}) \over 3 \beta_{0}}
 \log{1 \over \alpha_S(Q^2)}\,.
\label{eq:6.3.1}
\ee
Again, the decrease of $F_{2p}^{soft}(x,Q^{2})$ with $Q^{2}$ 
would only be possible at the expense of unphysical negative 
valued $G_{soft}(Q^2)$ at large $Q^2$.


\section{DGSF in the $x$-space: effective
intercepts and hard-to-soft diffusion}

\begin{figure}[!htb]
   \centering
   \epsfig{file=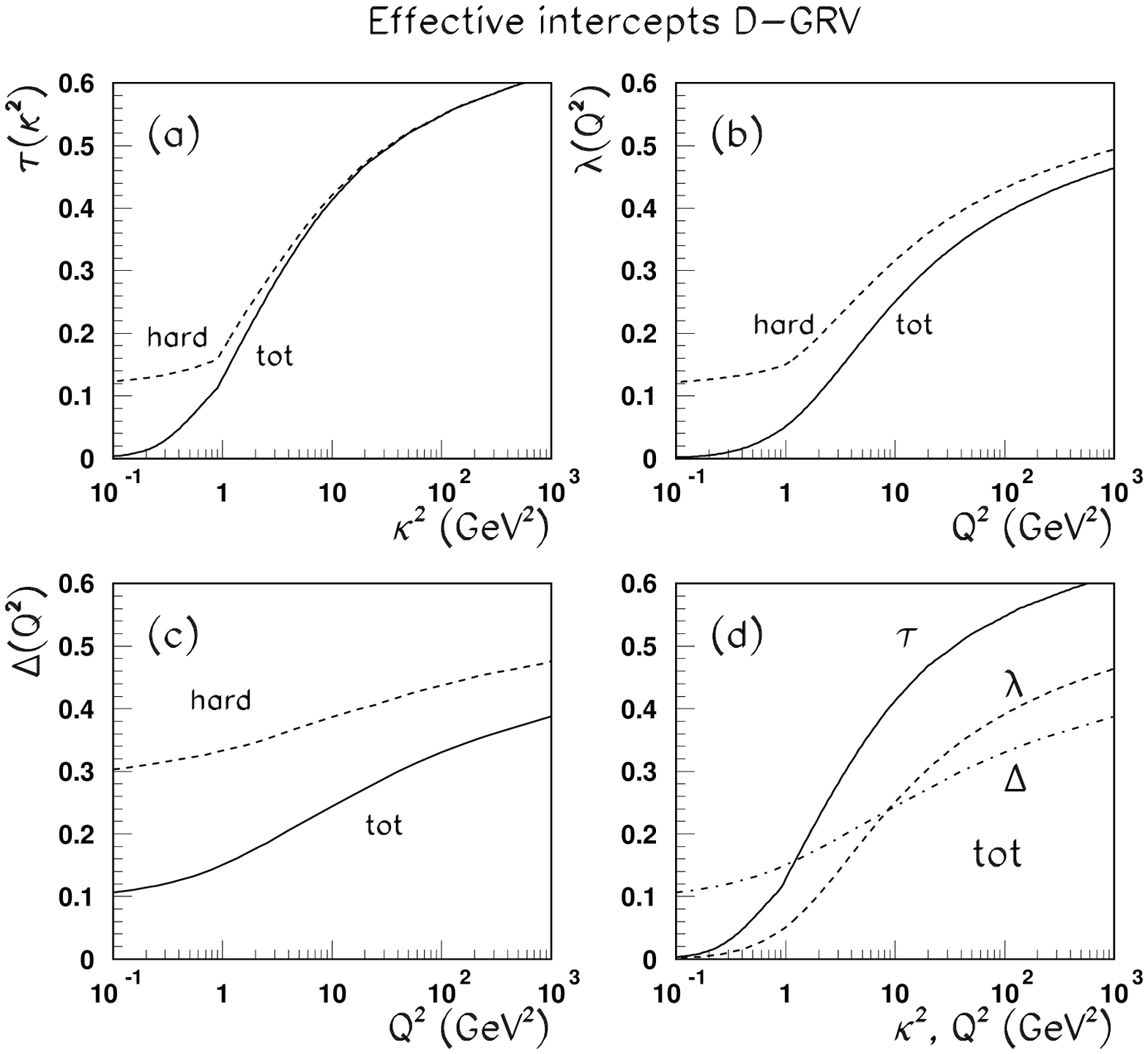,width=10cm}
   \caption{\em Effective intercepts for total, and
hard components of, (a) the differential
gluon structure function ${\cal F}(x,Q^{2})$; (b)
integrated gluon  structure function $G_{D}(x,Q^2)$ and
(c) 
proton structure function $F_{2p}(x,Q^{2})$ evaluated with the
D-GRV parameterization of the differential gluon structure 
function ${\cal F}(x,\bkappa^2)$. In the box (d) we compare 
the effective intercepts $\tau_{eff}(Q^2), \lambda_{eff}(Q^2)$
and $\Delta_{eff}(Q^2)$ for ${\cal F}(x,Q^{2})$, $G_{D}(x,Q^2)$
and $F_{2p}(x,Q^{2})$, respectively. }
   \label{InterceptsGRV}
\end{figure}

\begin{figure}[!htb]
   \centering
   \epsfig{file=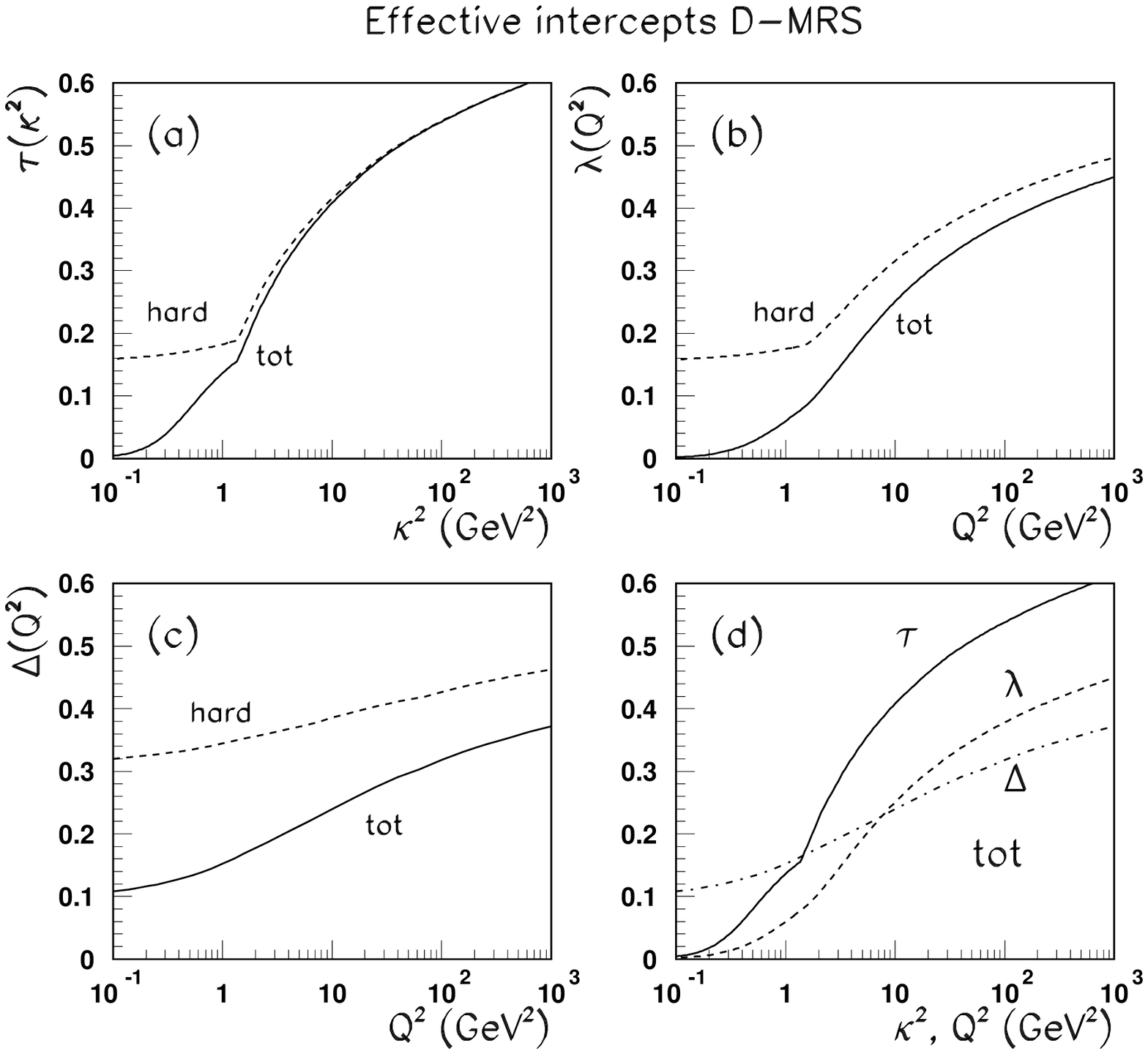,width=10cm}
   \caption{\em Effective intercepts for total, and
hard components of, (a)the differential
gluon structure function ${\cal F}(x,Q^{2})$; (b)  
integrated gluon  structure function $G_{D}(x,Q^2)$ and (c)
proton structure function $F_{2p}(x,Q^{2})$ evaluated with the
D-MRS parameterization of the differential gluon structure 
function ${\cal F}(x,\bkappa^2)$. In the box (d) we compare 
the effective intercepts $\tau_{eff}(Q^2), \lambda_{eff}(Q^2)$
and $\Delta_{eff}(Q^2)$ for ${\cal F}(x,Q^{2})$, $G_{D}(x,Q^2)$
and $F_{2p}(x,Q^{2})$, respectively.}
   \label{InterceptsMRS}
\end{figure}
\begin{figure}[!htb]
   \centering
   \epsfig{file=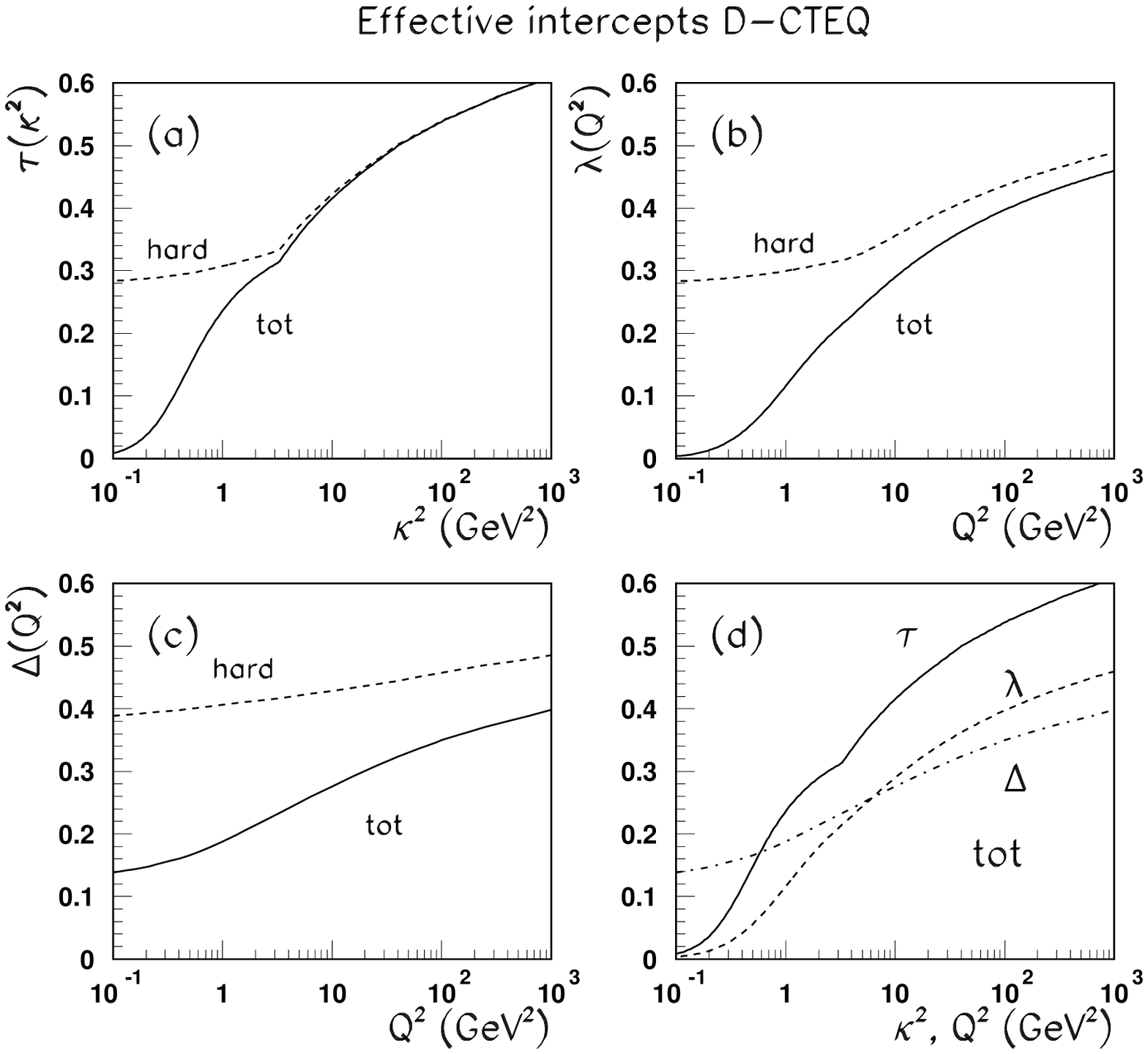,width=10cm}
   \caption{\em Effective intercepts for total, and
hard components of, (a) the differential
gluon structure function ${\cal F}(x,Q^{2})$; (b)  
integrated gluon  structure function $G_{D}(x,Q^2)$ and (c)
proton structure function $F_{2p}(x,Q^{2})$ evaluated with the
D-CTEQ parameterization of the differential gluon structure 
function ${\cal F}(x,\bkappa^2)$. In the box (d) we compare 
the effective intercepts $\tau_{eff}(Q^2), \lambda_{eff}(Q^2)$
and $\Delta_{eff}(Q^2)$ for ${\cal F}(x,Q^{2})$, $G_{D}(x,Q^2)$
and $F_{2p}(x,Q^{2})$, respectively.}
   \label{InterceptsCTEQ}
\end{figure}

It is instructive to look at the change of the $x$-dependence
from the differential gluon structure function ${\cal F}(x,Q^2)$
to integrated gluon structure function $G_{D}(x,Q^2)$ to proton
structure function $F_{2p}(x,Q^2)$.
It is customary to parameterize the $x$ dependence of various
structure functions by the effective intercept. For instance,
for the effective intercept $\tau_{eff}$ for differential gluon 
structure function is defined by the parameterization 
\be
{\cal F}(x,\bkappa^2) \propto \left({1 \over x}
\right)^{\tau_{eff}(\bkappa^2)}\,.
\label{eq:6.1}
\ee
One can define the related intercepts $\tau_{hard}$ for the 
hard component  ${\cal F}_{hard}(x,Q^2)$. Notice, that in
our Ansatz $\tau_{soft}\equiv 0$. 

The power law (\ref{eq:6.1}) is only a crude approximation to the 
actual $x$ dependence of DGSF and the effective intercept $\tau_{eff}$
will evidently depend on the range of fitted $x$. To be more
definitive, for the purposes of the present discussion we define 
the effective intercept as 
\be
\tau_{eff}(Q^2) = {\log [{\cal F}( x_{2},\bkappa^2)/
{\cal F}( x_{1},\bkappa^2)] \over 
\log(x_{1}/x_{2}) }
\label{eq:6.2}
\ee
taking ${x}_{2}=10^{-5}$ and ${x}_{1}=10^{-3}$. The effective intercept 
$\tau_{hard}(Q^2)$ is defined by (\ref{eq:6.2}) in terms of 
${\cal F}_{hard}(x,Q^2)$.

\begin{figure}[!htb]
   \centering
   \epsfig{file=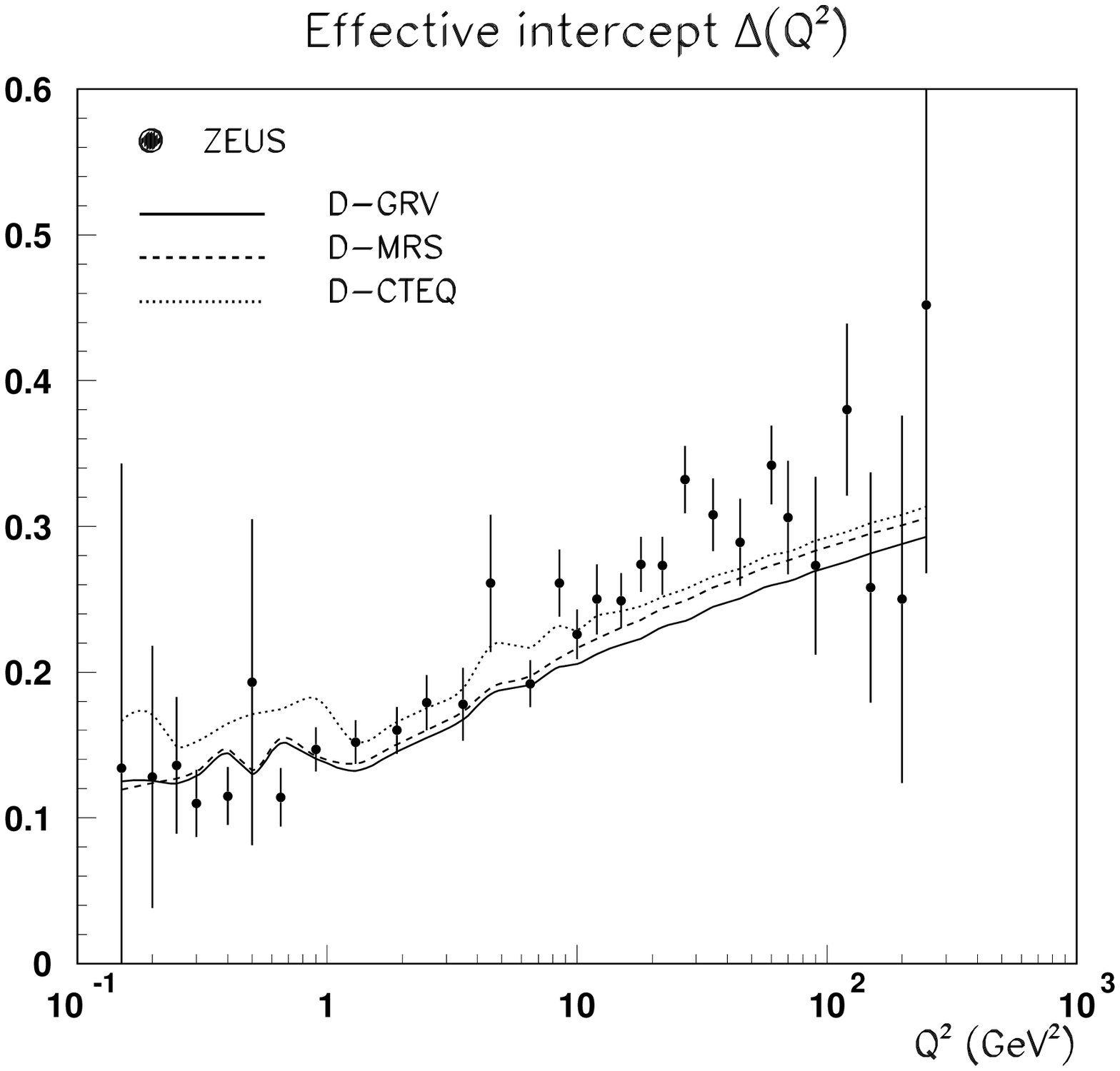,width=10cm}
   \caption{\em Effective intercepts $\Delta(Q^2)$ of the
proton structure function $F_{2p}(x,Q^{2})$ in the HERA domain
evaluated for the D-GRV, D-MRS and D-CTEQ parameterizations for 
the differential gluon structure 
function ${\cal F}(x,\bkappa^2)$;
the experimental data points are from ZEUS \cite{ZEUSshifted}}
   \label{Intercepts-HERA}
\end{figure}

One can define the related intercepts $\lambda_{eff},\lambda_{hard}$ 
for the integrated gluon structure function $G_{D}(x,Q^2)$:
\be
G_{D}(x,Q^2) \propto \left({1 \over x}
\right)^{\lambda_{eff}(Q^2)}\,.
\label{eq:6.3}
\ee

In the case of $F_{2p}(x,Q^2)$ we define the intercept $\Delta(Q^2)$
in terms of the variable ${\overline x}$ defined as
\be
{\bar x} = {Q^2 +M_{V}^2 \over W^2 +Q^2}\sim x_g\, ,
\label{eq:5.2.0}
\ee
where $M_{V}$ is the mass of the ground state vector meson 
in the considered flavor channel. Such a replacement allows one
to treat on equal footing $Q^{2} \lsim 1$ GeV$^2$,
where the formally defined Bjorken variable $x_{bj}$ 
can no longer be interpreted as a lightcone momentum 
carried by charged partons.  For the purposes of the direct comparison with
$\tau(Q^2),\lambda(Q^2)$ and in 
order to avoid biases caused by the valence structure function,
here we focus on intercepts  $\Delta_{eff},\Delta_{hard}$  for the 
sea component of the proton structure function $F_{2p}^{sea}(x,Q^2)$:
\be
F_{2p}^{sea}(x,Q^2) \propto \left({1 \over {\overline x}}
\right)^{\Delta_{eff}(Q^2)}\,.
\label{eq:6.4}
\ee
The results for the effective intercepts are shown in 
figs.~\ref{InterceptsGRV}, \ref{InterceptsMRS} and \ref{InterceptsCTEQ}.

In our simplified hard-to-soft extrapolation of ${\cal F}_{hard}(x,Q^2)$
we attribute to ${\cal F}_{hard}(x,Q^2)$ at $Q^2 \leq Q_{c}^{2}$ the same 
$x$-dependence as at $Q^{2}=Q_c^2$ modulo to slight modifications for
the $x$-dependence of $\bkappa_h^2$. This
gives the cusp in $\tau_{hard}(Q^{2})$ at $Q^2=Q_{c}^2$, i.e., the
first derivative of $\tau_{hard}(Q^{2})$ is discontinuous at $Q^2=Q_{c}^2$.

A comparison of fig.~\ref{DGSFoverlaid} with 
fig.~\ref{GSF} and further with fig.~\ref{Soft-Hard.F2p} shows clearly 
that only in DGSF ${\cal F}(x,Q^2)$ the effect of the soft component
is concentrated at small $Q^{2}$, in integrated $G_{D}(x,Q^2)$ and especially
in the proton structure function $F_{2p}(x,Q^2)$ the impact of
the soft component extends to much larger $Q^{2}$. The larger
the soft contribution, the stronger is the reduction of $\tau_{eff}$ 
from $\tau_{hard}$ and so forth, the pattern which is evident from
fig.~\ref{InterceptsGRV}a to \ref{InterceptsGRV}b to \ref{InterceptsGRV}c,
see also figs.~\ref{InterceptsMRS} and \ref{InterceptsCTEQ}.

The change of effective intercepts from differential ${\cal F}(x,Q^2)$
to integrated $G_{D}(x,Q^2)$ is straightforward, the principal effect is 
that $\lambda_{hard}(Q^{2}) < \tau_{hard}(Q^{2})$ and  
$\lambda_{eff}(Q^{2}) < \tau_{eff}(Q^{2})$ which reflects the growing
importance of soft component in $G_{D}(x,Q^2)$.
The change of effective intercepts from ${\cal F}(x,Q^2)$ and $G_{D}(x,Q^2)$  
to  $ F_{2p} (x,Q^2)$ is less trivial and exhibits two dramatic consequences  
of the hard-to-soft and soft-to-hard diffusion. If the standard DGLAP 
contribution (\ref{eq:3.2.2}) were all, 
then the change from the intercept $\lambda(Q^2)$ for integrated
gluon density to the intercept $\Delta(Q^{2})$ for the proton structure 
function $ F_{2p} (x,Q^2)$ would have been similar to the
change from $\tau(Q^2)$ to $\lambda(Q^2)$, i.e., the effective intercept
$\Delta_{eff}(Q^2)$ would have been close to zero for $Q^2 \lsim 1$ GeV$^2$.
However, by virtue of the hard-to-soft diffusion phenomenon inherent to the 
$\bkappa$-factorization, $ F_{2p}(x,Q^2)$ receives a contribution 
from gluons with $\bkappa^2 > Q^2$, which enhances substantially 
$\Delta_{hard}(Q^{2})$ and $\Delta_{eff}(Q^{2})$.
The net result is that at small to moderately large  
$Q^{2}$ we find $\Delta_{hard}(Q^{2})> \lambda_{hard}(Q^{2})$
and $\Delta_{eff}(Q^{2})> \lambda_{eff}(Q^{2})$.  As we emphasized 
above in sections 5.3, the rise of real photoabsorption cross
section is precisely of the same origin. 

The second effect is a dramatic flattening of effective hard intercept, 
$\Delta_{hard}(Q^{2})$, over the whole range of $Q^{2}$. For all 
three DGLAP inputs $\Delta_{hard}(Q^{2})$ flattens at approximately 
the same $\Delta_{hard} \approx 0.4$.

The whole set of figs.~\ref{InterceptsGRV}--\ref{InterceptsCTEQ} 
also shows that the systematics of intercepts 
in the hard region of $Q^{2} > Q_{c}^{2}$ 
is nearly identical for all the three DGLAP inputs.
In the soft region we have a slight inequality $\left.
\tau_{hard}(\bkappa^{2})\right|_{D-MRS} >  \left.\tau_{hard}(\bkappa^{2})
\right|_{D-GRV}$, which can be readily attributed to a slight inequality
$Q_{c}^2(MRS)  > Q_c^2(GRV)$. In the case of  CTEQ4L(v.4.6) input 
the value of $Q_{c}^{2}(CTEQ)$ is substantially larger then
$Q_{c}^2(MRS), Q_c^2(GRV)$. In the  
range  $Q_{c}^2(MRS), Q_c^2(GRV) < \bkappa^{2}
< Q_{c}^{2}(CTEQ)$ the effective intercept $\tau_{hard}(\bkappa^2)$ 
rises steeply with $\bkappa^2$. This explains a
why in the soft region $\left. \tau_{hard}(\bkappa^{2})\right|_{CTEQ}$ 
is significantly larger than for the D-GRV and D-MRS parameterizations. The
difference among intercepts for the three parameterizations decreases
gradually from differential ${\cal F}(x,\bkappa^2)$ to integrated 
$G_{D}(x,Q^2)$ gluon density to the proton structure function $F_{2p}(x,Q^2)$.

Finally, in fig.~\ref{Intercepts-HERA} we compare our results 
for $\Delta_{eff}(Q^2)$ with the recent experimental data 
from ZEUS collaboration \cite{ZEUSshifted}. 
In the experimental fit the range of $x=[x_{max},x_{min}]$  varies from point 
to point, in our evaluation of $\Delta_{eff}$ from
eq.~(\ref{eq:7.4}) we mimicked the experimental 
procedure taking ${\overline x}_{2}=x_{max}$ and ${\overline x}_{1}=x_{min}$.
This explains the somewhat irregular $Q^{2}$ dependence. 
The experimental data include both sea and valence components.
At $Q^2 > Q_{c}^{2}(GRV) =0.9$ GeV$^2$ we included the valence 
component of the structure function taking the GRV98LO parameterization. 
For CTEQ4L(v.4.6) and MRS-LO-1998 
the values of $Q_{c}^2$ are substantially larger. However, the valence 
component is a small correction and we took a liberty of evaluating
the valence contribution $F_{2p}^{val}(x,Q^2)$ for $ Q_{c}^{2}(GRV)<Q^2< 
Q_{c}^{2}(MRS), Q_{c}^{2}(CTEQ)$. The overall agreement
with experiment is good.
Difference among the three parameterization is marginal and 
can of course be traced back to 
figs.~\ref{InterceptsGRV}--\ref{InterceptsCTEQ}.


\section{How the gluon densities of $\bkappa$-factorization 
differ from DGLAP gluon densities}

\begin{figure}[!htb]
   \centering
   \epsfig{file=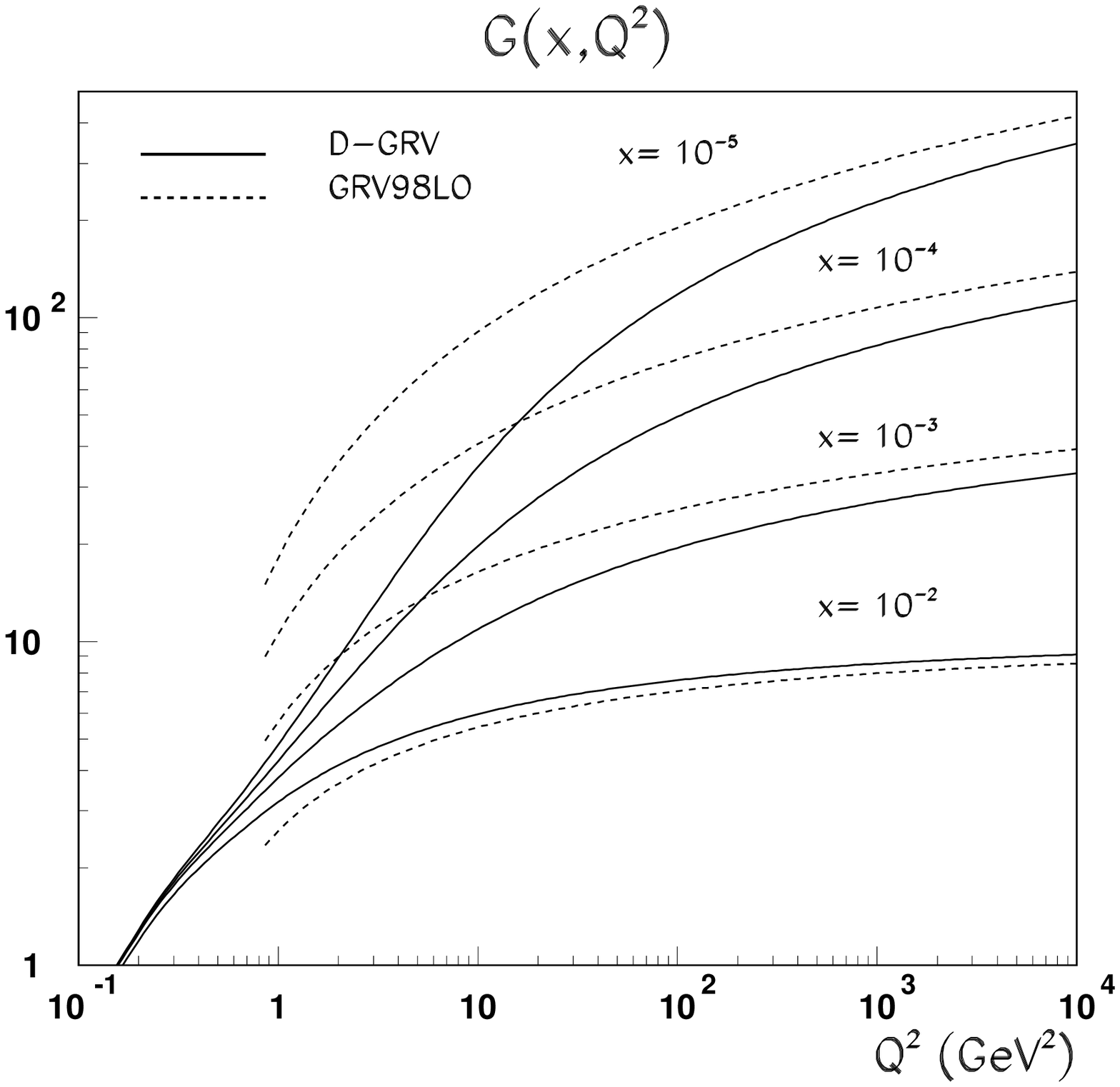,width=10cm}
   \caption{\em Comparison of our results for integrated gluon density
$G_{D}(x,Q^{2})$ evaluated with the
D-GRV parameterization of the differential gluon structure 
function ${\cal F}(x,\bkappa^2)$  with the GRV98L0 DGLAP input parameterization
$G_{pt}(x,Q^2)$. }
   \label{D-GRV.vs.GRV}
\end{figure}

\begin{figure}[!htb]
   \centering
   \epsfig{file=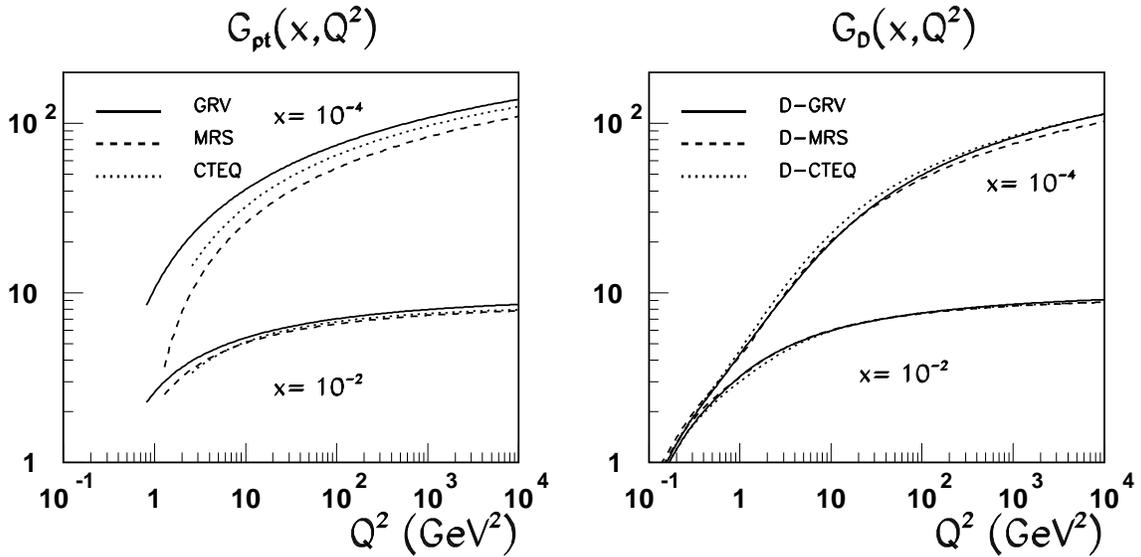,width=15cm}
   \caption{\em A comparison of the divergence of 
GRV98L0, CTEQ4L(v.4.6) and MRS-LO-1998 gluon
structure functions $G_{pt}(x,Q^2)$ in the left box 
with the divergence of our integrated gluon structure 
functions $G_{D}(x,Q^2)$ evaluated for the
D-GRV, D-CTEQ and D-MRS parameterizations for
differential gluon structure function ${\cal F}(x,Q^2)$
at two typical values of $x$}
   \label{G_input.vs.G}
\end{figure}

\begin{figure}[!htb]
   \centering
   \epsfig{file=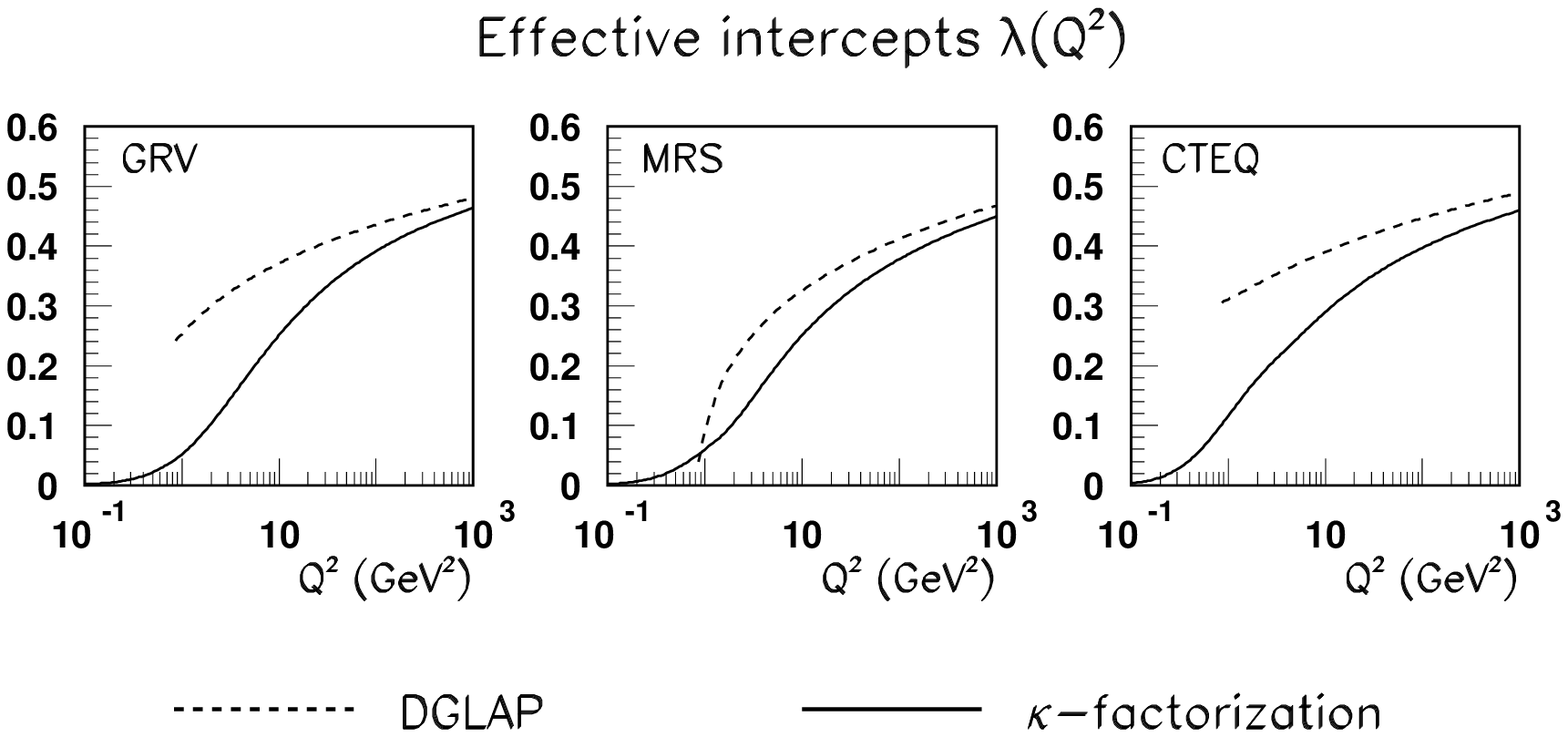,width=16cm}
   \caption{\em A comparison of the intercept $\lambda_{eff}^{(pt)}(Q^2)$
of the $x$-dependence of the GRV98L0, CTEQ4L(v.4.6) and MRS-LO-1998 gluon
structure functions $G_{pt}(x,Q^2)$  with their  counterpart 
$\lambda_{eff}(Q^2)$ for integrated $G_{D}(x,Q^2)$ evaluated 
with D-GRV, D-CTEQ and D-MRS parameterizations for
differential gluon structure function ${\cal F}(x,Q^2)$.}
   \label{Intercepts-DGLAP.vs.Differ}
\end{figure}

It is instructive also to compare our results for integrated
GSF (\ref{eq:5.2.1}) with the conventional DGLAP fit $G_{pt}(x,Q^2)$.
In fig.~\ref{D-GRV.vs.GRV} we present such a comparison
 between our integrated D-GRV distribution  
(the solid curves) and the GRV98LO distribution
(the dashed curves). As was anticipated in section 3.2,  
at very large $Q^{2}$ the two gluon distributions converge. 
We also anticipated that at small $x$ and moderate $Q^{2}$ the DGLAP
gluon structure functions $G_{pt}(x,Q^2)$ are substantially larger 
than the result of integration of DGSF, see eq.~(\ref{eq:5.2.1}).  
At $x=10^{-5}$ they differ by as much as
the factor two-three over a broad range of $Q^{2} \lsim 100$ GeV$^{2}$.
The difference between integrated DGSF and the DGLAP fit decreases 
gradually at large $x$, and is only marginal at $x=10^{-2}$.

Recall the substantial divergence of the GRV, MRS and CTEQ gluons 
structure functions of DGLAP approximation 
$G_{pt}(x,Q^2)$ at small and moderate $Q^{2}$. Contrary to 
that, the   $\bkappa$-factorization D-GRV, D-CTEQ and D-MRS gluon 
structure functions $G_D(x,Q^2)$ are 
nearly identical. We demonstrate this property 
in fig.~\ref{G_input.vs.G} where we 
show integrated $G_D(x,Q^2)$ and their DGLAP counterparts $G_{pt}(x,Q^2)$ 
for the three parameterizations at two typical
values of $x$. Because of an essentially unified treatment of
the region of $\bkappa^{2} \leq Q_{c}^{2}$ and strong constraint
on DGSF in this region from the experimental data at small $Q^{2}$,
such a convergence of D-GRV, D-CTEQ and D-MRS DGSF's is not
unexpected.

One can also compare the effective intercepts for our integrated GSF
$G_{D}(x,Q^2)$ with those obtained from DGLAP gluon distributions 
$G_{pt}(x,Q^2)$. Fig.\ref{Intercepts-DGLAP.vs.Differ} shows large 
scattering of $\lambda^{(pt)}_{eff}(Q^2)$ from one DGLAP input to another.
At the same time, this divergence of different DGLAP input
parameterizations 
is washed out to a large extent 
in the $\bkappa$-factorization description of
physical observables (see also \ref{Intercepts-HERA}).


\section{How different observables probe the DGSF ${\cal F}(x,Q^{2})$}

The issue we address in this section 
is how different observables map the $\bkappa^2$ dependence of 
${\cal F}(x_{g},\bkappa^{2})$. 
We expand on the qualitative discussion in section 3.2 and 
corroborate  it with numerical analysis following the discussion 
in \cite{NZglue}. . We start
with the two closely related quantities ---  longitudinal
structure function $F_{L}(x,Q^{2})$ and scaling violations
$\partial F_{2}(x,Q^{2})/\partial \log Q^{2}$ --- and proceed to
$F_{2p}(x,Q^{2})$ and the charm structure function of the proton
$F_{2p}^{c\bar{c}}(x,Q^{2})$.  This mapping is best studied if in
(\ref{eq:3.1.7}) and (\ref{eq:3.1.8}) we integrate first over
$\bk$ and $z$. In order to focus on the $\bkappa^2$ 
dependence we prefer presenting different observables in terms of  
${\cal F}(2x,\bkappa^{2})$ and $G_{D}(2x,\bkappa^{2})$ 
\bea
F_{L}(x,Q^{2}) = 
{\alpha_{S}(Q^{2}) \over 3\pi} \sum e_{f}^{2}
\int {d\bkappa^{2}\over \bkappa^{2}}\Theta_{L}^{(f\bar{f})}(Q^{2},\bkappa^{2})
{\cal F}(2x,\bkappa^{2})\, ,
\label{eq:7.1}
\eea
\bea
{\partial F_{2}(x,Q^{2})\over \partial \log Q^{2}} =
{\alpha_{S}(Q^{2}) \over 3\pi} \sum e_{f}^{2}
\int {d\bkappa^{2}\over \bkappa^{2}}\Theta_{2}^{(f\bar{f})}(Q^{2},\bkappa^{2})
{\cal F}(2x,\bkappa^{2})\, .
\label{eq:7.2}
\eea
\begin{figure}[!htb]
   \centering
   \epsfig{file=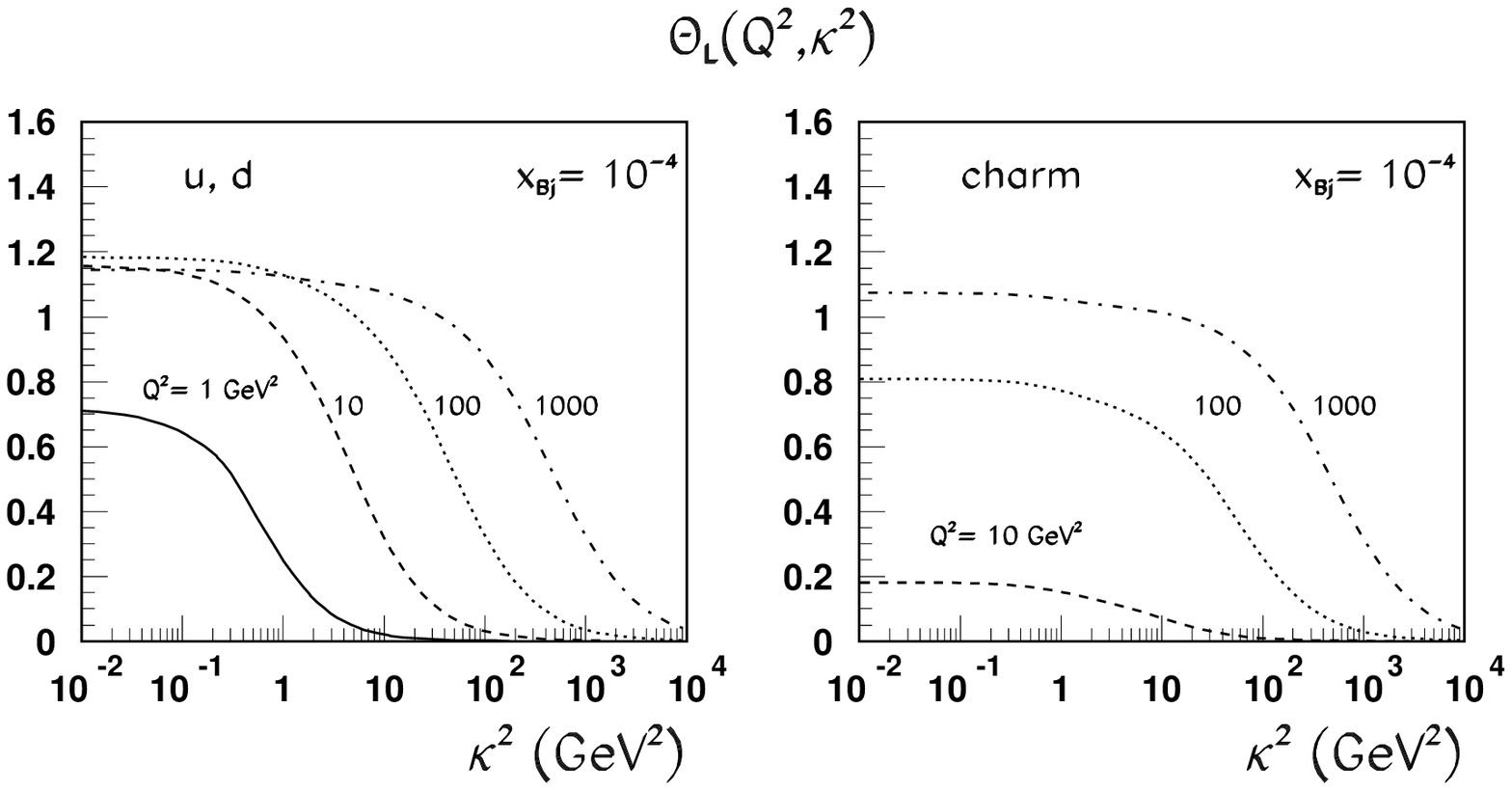,width=16cm}
   \caption{\em The weight function $\Theta_{L}$ for mapping 
of the differential gluon structure function ${\cal F}(x,\bkappa^{2})$ as a function of
$\bkappa^{2}$ for several values of $Q^{2}$. We show separately the results
for
light flavours, $u,d$, and charm. }
   \label{ThetaL}
\end{figure}
 
\begin{figure}[!htb]
   \centering
   \epsfig{file=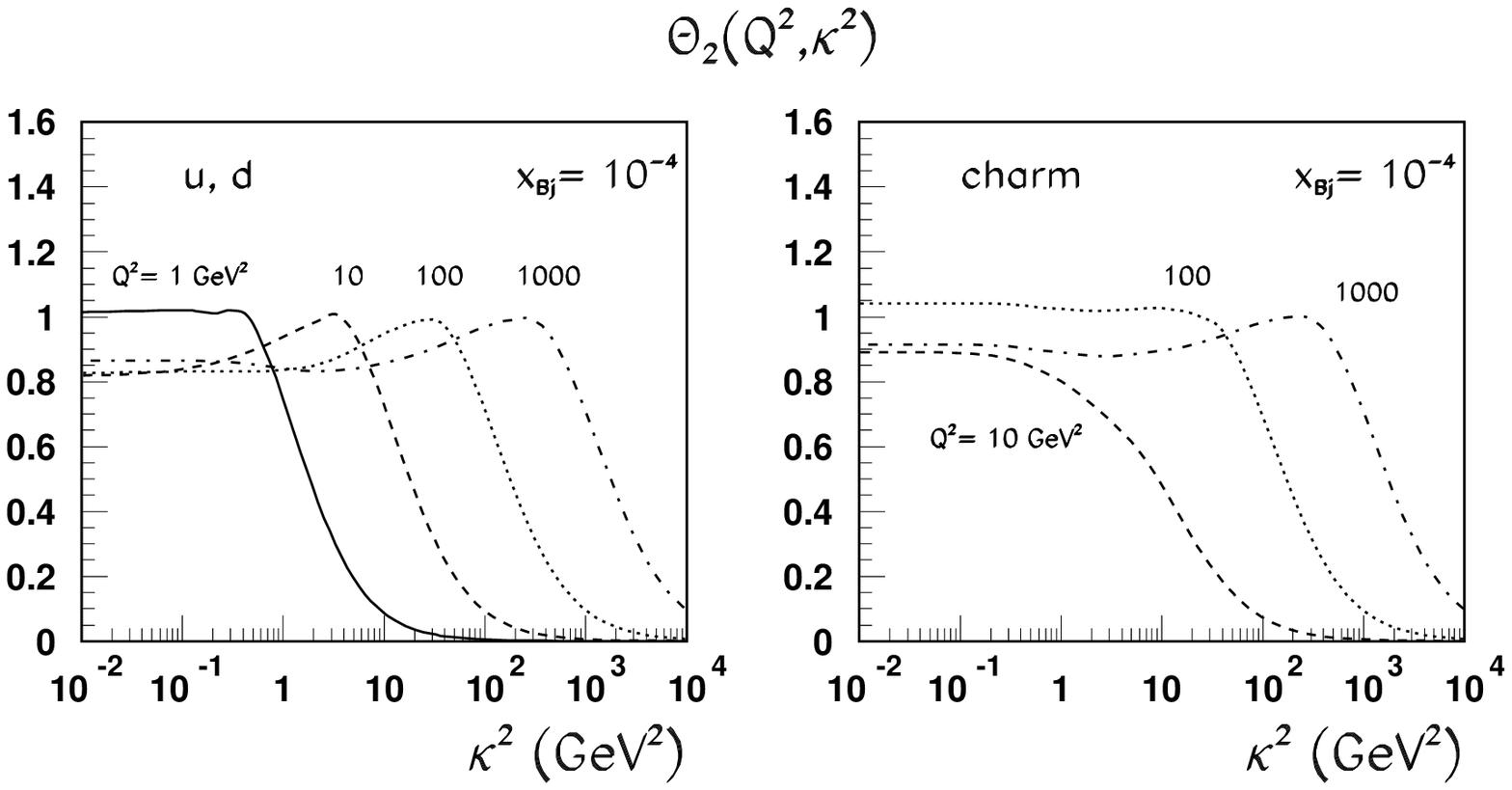,width=16cm}
   \caption{\em The weight function $\Theta_{2}$ for mapping 
of the differential gluons structure function ${\cal F}(x,\bkappa^{2})$
as a function of
$\bkappa^{2}$ for several values of $Q^{2}$. We show separately the results
for
light flavours, $u,d$, and charm.}
   \label{Theta2}
\end{figure}

In the numerical calculation of $F_{L}(x,Q^{2})$ starting from 
eq.~(\ref{eq:3.1.8}) we have $x_{g}$ and $\bkappa^{2}$ 
as the two running arguments of  ${\cal F}(x_{g},\bkappa^{2})$. 
As discussed above, the mean value of $x_{g}$ is close to $2x$, but the exact 
relationship depends on $\bkappa^2$.  The $\bk,z$ integration
amounts to averaging of ${\cal F}(x_{g},\bkappa^{2})$ over certain 
range of $x_{g}$. The result of this averaging is for the most part
controlled by the effective intercept $\tau_{eff}(\bkappa^2)$: 
\be
\langle {\cal F}(x_{g},\bkappa^{2})\rangle =
\left\langle {\cal F}(2x,\bkappa^{2})
\left({2x \over x_{g}}\right)^{\tau_{eff}(\bkappa^{2})}\right\rangle
=r(\bkappa^2) {\cal F}(2x,\bkappa^{2})\, .
\label{eq:7.3}
\ee
Because the derivative of $\tau_{eff}(\bkappa^2)$ changes rapidly around
$\bkappa^2=Q_{c}^2$, the rescaling factor $r(\bkappa^2)$ also has 
a rapid variation of the  derivative at $\bkappa^2=Q_{c}^2$,
which in the due turn generates the rapid change of derivatives of 
$\Theta_{L,2}^{(f\bar{f})}(Q^{2},\bkappa^{2})$ around $\bkappa^2=Q_{c}^2$.
As far as the mapping of differential ${\cal F}(2x,\bkappa^{2})$ 
is concerned, this is an entirely marginal effect. 
However, if we look at the mapping of integrated gluon structure 
function $G_{D}(x,Q^{2})$, which is derived from (\ref{eq:7.1}), 
(\ref{eq:7.2}) by integration by parts:
\bea
F_{L}(x,Q^{2}) = 
-{\alpha_{S}(Q^{2}) \over 3\pi} \sum e_{f}^{2}
\int {d\bkappa^{2}\over \bkappa^{2}}{\partial \Theta_{L}^{(f\bar{f})}(Q^{2},\bkappa^{2})
\over \partial \log\bkappa^2}
G_{D}(2x,\bkappa^{2})
\, ,
\label{eq:7.4}
\eea
\bea
{\partial F_{2}(x,Q^{2})\over \partial \log Q^{2}} =
-{\alpha_{S}(Q^{2}) \over 3\pi} \sum e_{f}^{2}
\int {d\bkappa^{2}\over \bkappa^{2}}{\partial \Theta_{2}^{(f\bar{f})}(Q^{2},\bkappa^{2})
\over \partial \log\bkappa^2}
G_{D}(2x,\bkappa^{2})\, ,
\label{eq:7.5}
\eea
then the weight functions $\partial \Theta_{2,L}^{(f\bar{f})}
(Q^{2},\bkappa^{2}) / \partial \log\bkappa^2$ 
will exhibit a slightly irregular behaviour around
$\bkappa^2=Q_{c}^2$. Evidently, such an irregularity appears in any region of
fast variation of $\tau_{eff}(\bkappa^2)$;
in our simplified model it is somewhat amplified
by the cusp-like $\bkappa^2$ dependence of $\tau_{eff}(\bkappa^2)$.

\begin{figure}[!htb]
   \centering
   \epsfig{file=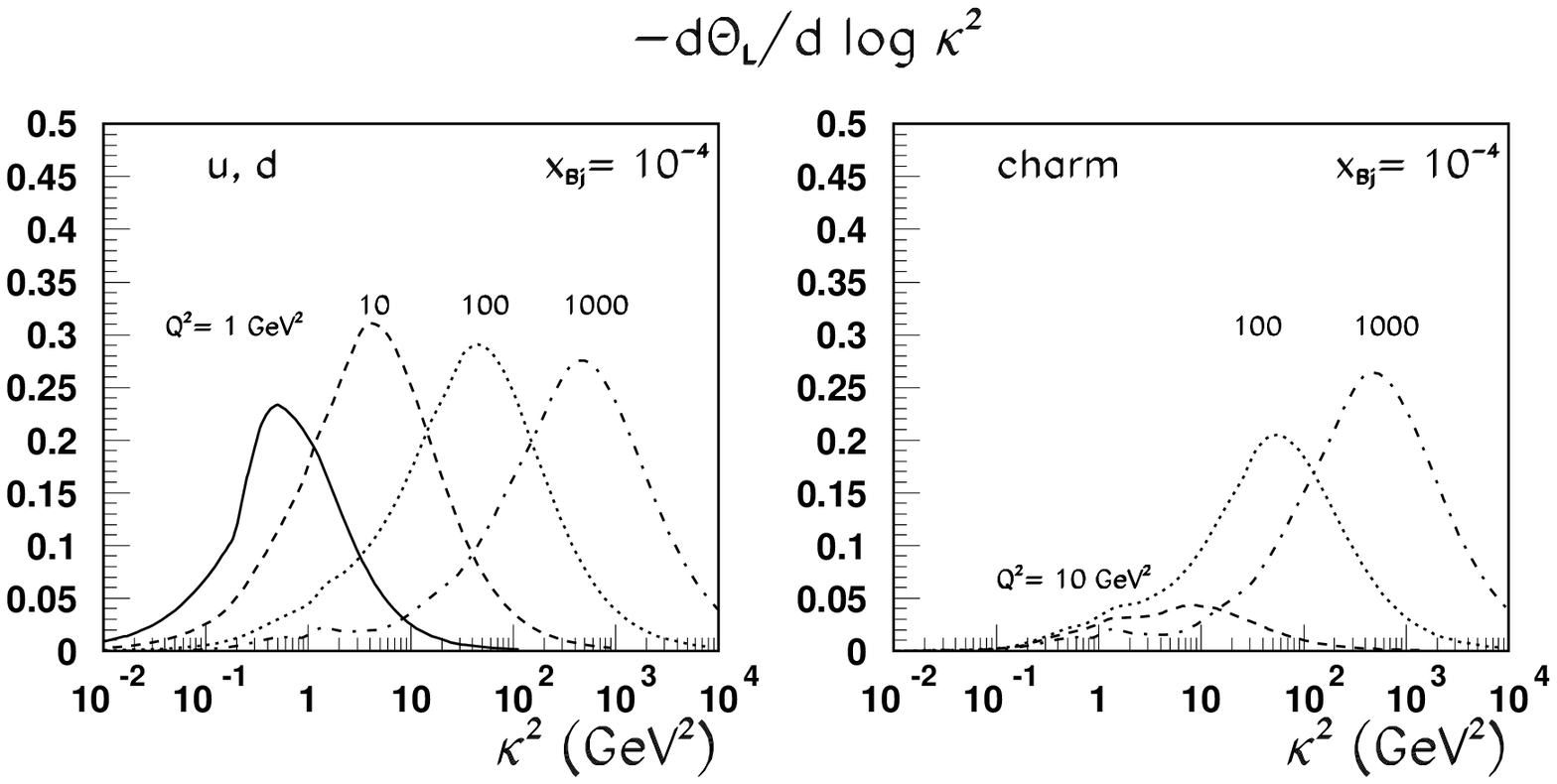,width=16cm}
   \caption{\em The same as Fig.~\ref{ThetaL} but for mapping 
of the integrated gluon structure function $G_{D}(x,\bkappa^{2})$ 
as a function of $\bkappa^{2}$ for several values of $Q^{2}$. 
We show separately the results for light flavours and charm. }
   \label{PeakL}
\end{figure}

Finally, starting from (\ref{eq:7.5}) one obtains a useful representation
for how the proton structure function $F_{2p}(x,Q^2)$ maps the integrated
gluon structure function: 
\bea
F_{2p}(x,Q^{2})=-\int_{0}^{Q^{2}}{dq^2 \over q^{2}}
{\alpha_{S}(q^{2}) \over 3\pi} \sum e_{f}^{2}
\int {d\bkappa^{2}\over \bkappa^{2}}{\partial \Theta_{2}^{(f\bar{f})}(q^{2},\bkappa^{2})
\over \partial \log\bkappa^2}
G_{D}(2x,\bkappa^{2}) \nonumber\\
={1\over 3\pi} \sum e_{f}^{2} \int {d\bkappa^{2}\over \bkappa^{2}}W^{(f\bar f)}_{2}(Q^{2},\bkappa^{2})\alpha_{S}(\bkappa^2)
G_{D}(2x,\bkappa^{2})
\label{eq:7.6}
\eea
 
\begin{figure}[!htb]
   \centering
   \epsfig{file=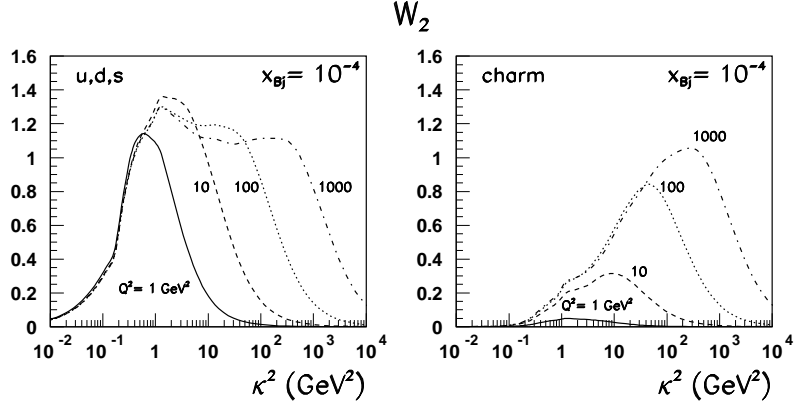,width=12cm}
   \caption{\em The weight function $W_{2}$ for mapping 
of the integrated gluons structure function $G_{D}(x,\bkappa^{2})$ 
as a function of $\bkappa^{2}$ for several values of $Q^{2}$. 
We show separately the results for light flavours and charm}
   \label{W2}
\end{figure}

\begin{figure}[!htb]
   \centering
   \epsfig{file=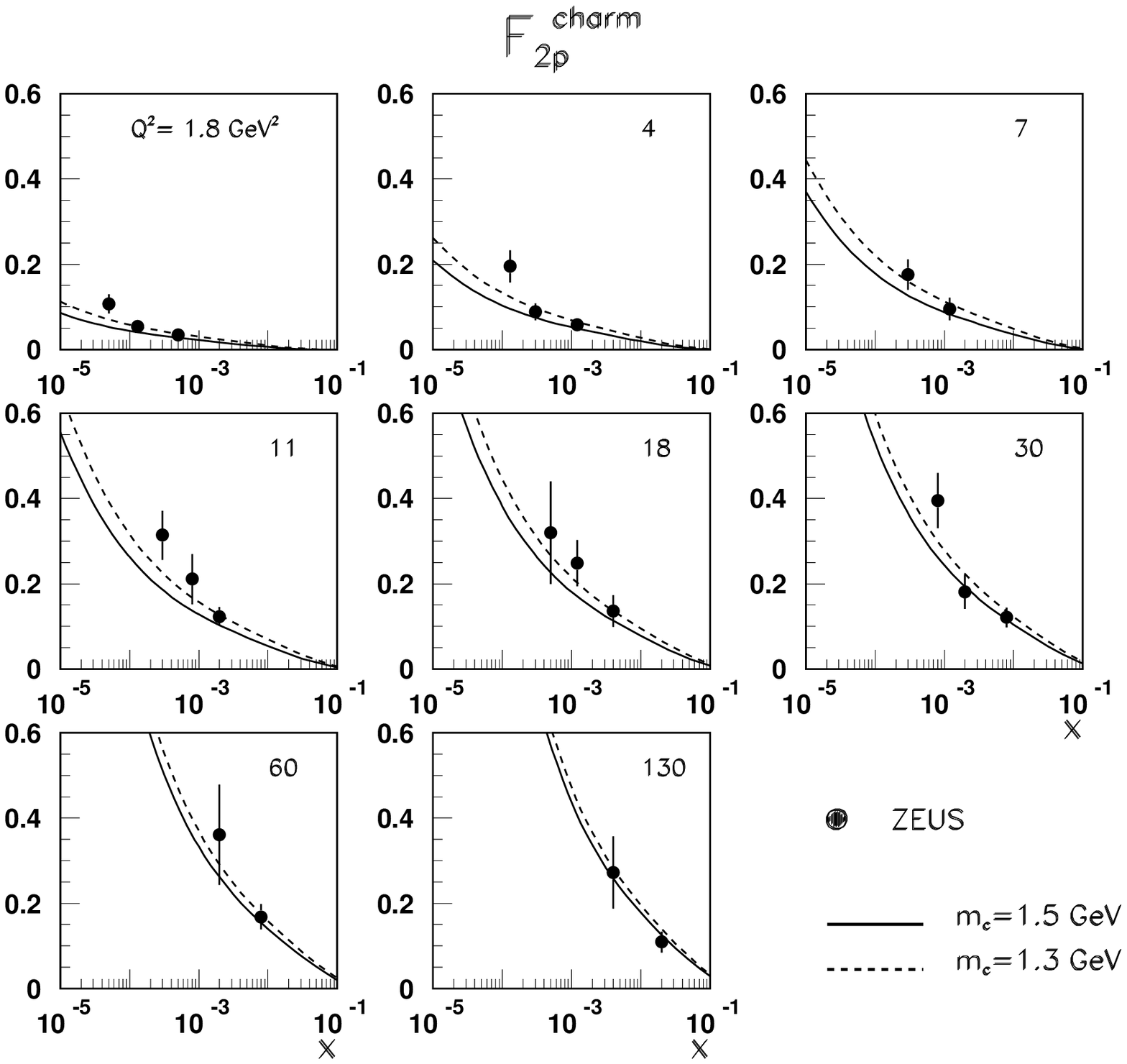,width=12cm}
   \caption{\em A comparison of the  experimental data from ZEUS \cite{ZEUScharm}
on the charm structure function of the proton with $\bkappa$-factorization 
results for $F_{2}^{c\bar{c}}(x,Q^2)$ based 
on the D-GRV parameterization of the differential gluon
structure function ${\cal F}(x,Q^2)$.}
   \label{F2charm}
\end{figure}

In figs.~\ref{ThetaL} and \ref{Theta2} we show the weight functions 
$\Theta_{L}$ and $\Theta_{2}$.
Evidently, for light flavours and very large $Q^{2}$ they can be approximated 
by step-functions
\be
\Theta_{L,2}^{(f\bar{f})}(Q^{2},k^{2})\sim \theta(C_{L,2}Q^{2}-\bkappa^2)\,,
\label{eq:7.7}
\ee
where the scale factors $C_{L} \sim {1\over 2}$ and $C_{2} \sim 2$ can
be readily read from figures, for the related discussion see \cite{NZglue}. 
Note that the value $C_2\sim 2$ corresponds to $\bar C_2 \sim 8$
introduced in Section 3.2.
Recall that the development of the plateau-like behaviour of $\Theta_{L}$ and 
$\Theta_{2}$ which extends to $\bkappa^{2}\sim Q^{2}$ signals the onset of 
the leading log$Q^{2}$ approximation. For large $Q^{2}$ in the approximation
(\ref{eq:7.7}) the $\bkappa^2$ integration can be carried out explicitly and
$F_{L}(x,Q^{2}) \propto  G_{D}(2x,C_{L}Q^2)$. Similarly, $
\partial F_{2}(x,Q^{2})/\partial \log Q^{2} \propto  G_{D}(2x,C_{2}Q^2)$, cf.
eq.~(\ref{eq:3.2.6}).

Still better idea on how $F_{L}$ and scaling violations map the integrated
GSF is given by figs.~\ref{PeakL},\ref{W2}, where we show results for
$-\partial \Theta{(f\bar f)}_L/\partial \log\bkappa^2$ and $W^{(f\bar f)}_2$.
The first quantity is sharply peaked at $\kappa^2 \sim C_{L}Q^{2}$. 
The second quantity visibly develops a plateau at large $Q^2$.
As can be easily seen, scaling violations do receive a substantial contribution
from the beyond-DGLAP region of $\bkappa^{2}> Q^{2}$. 

Because of the heavy mass, the case of the charm structure function 
$F_{2p}^{c\bar{c}}(x,Q^{2})$ is somewhat special. 
Figs.~\ref{PeakL} and \ref{W2} show weak sensitivity of 
$F_{2p}^{c\bar{c}}(x,Q^{2})$ to a soft component of 
${\cal F}(x_{bj},\bkappa^{2})$ which has an obvious
origin: long-wavelength soft gluons with $\kappa \lsim m_{c}$ decouple from 
the color neutral $c\bar{c}$ Fock state  of the 
photon which has a small transverse size $\lsim {1\over m_{c}}$. Our results
for $F_{2p}^{c\bar{c}}(x,Q^{2})$ are shown in fig.~\ref{F2charm}, 
the agreement with the recent precision experimental data 
from ZEUS \cite{ZEUScharm} is good.
  

\section{Summary and outlook}

We present the first parameterization of differential gluon
structure function ${\cal F}(x,Q^2)$ of the proton inherent
to the $\bkappa$-factorization approach to small-$x$ DIS. The form of the 
parameterization is driven by color gauge invariance constraints 
for soft $Q^2$, early ideas from color dipole phenomenology on
the necessity of nonperturbative soft mechanism for interaction
of large color dipoles  and by matching to the derivative of familiar 
DGLAP fits $G_{pt}(x,Q^2)$. The latter condition is not imperative,
though, and can be relaxed; in this exploratory study we simply 
wanted to take advantage of the insight on $G_{pt}(x,Q^2)$
from early DGLAP approximation studies on scaling violations.  The parameters
of ${\cal F}(x,Q^2)$ have been tuned to the experimental data on 
$F_{2p}$ in the low-$x$
($x \lsim 0.01$) domain and throughout the entire $Q^2$ region
as well as on real photoabsorption cross section 
$\sigma^{\gamma p}_{tot}$. Differential gluon structure function 
${\cal F}(x,Q^2)$ is the principal input for pQCD calculation of
many diffractive processes and we anticipate that the consistent
use of our parameterizations shall reduce the uncertainties of
calculations of cross section of such processes as diffractive DIS 
into vector mesons and continuum.

Our results allow to address several interesting issues. First,
our Ans\"atze for ${\cal F}(x,Q^2)$ have been so constructed as
to ensure the convergence of $G_{D}(x,Q^2)$ --- the integral of
${\cal F}(x,Q^2)$ --- to the corresponding large $Q^2$  DGLAP input
$G_{pt}(x,Q^2)$. We notice that both gluon distributions provide
a comparable description of the same set of the experimental data
on the proton structure function, the only difference being that
in the $\bkappa$-factorization we lift the DGLAP limitation on 
the transverse phase space of quarks and antiquarks. We find
very slow convergence of, and numerically very large difference
between, the $\bkappa$-factorization distribution 
$G_{D}(x,Q^2)$ and the DGLAP fit $G_{pt}(x,Q^2)$. As anticipated,
the divergence of the two distributions is especially
large at small-$x$ and persists even in the hard region up to 
$Q^{2}\sim 10\div 100$ GeV$^2$ at $x=10^{-5}$. We interpret this
divergence as a signal of breaking of the DGLAP approximation
which arguably gets poorer at smaller $x$. 
The second finding is a numerically very strong
impact of soft gluons on the integrated gluons structure function
$G_{D}(x,Q^2)$ and the proton structure function $F_{2p}(x,Q^2)$.
It is not unexpected in view of the early work on color dipole
phenomenology of small-$x$ DIS, but the evaluation of the soft 
component of integrated gluon structure function is reported here 
for the first time. In conjunction with the strong departure
of the $\bkappa$-factorization distribution 
$G_{D}(x,Q^2)$ from the DGLAP fit $G_{pt}(x,Q^2)$ it serves as an 
warning against unwarranted application of DGLAP evolution at
$Q^{2}$ in the range of several GeV$^2$.  

The phenomenologically most interesting finding is the anatomy
of the rising component of the proton structure function from
the Regge theory point of view. We notice that effective intercepts
$\tau_{hard}(Q^2)$ and $\lambda_{hard}$ for hard components of the 
differential and integrated gluon distributions are lively functions
of $Q^2$ which vary quite rapidly with $Q^2$ from $\approx 0.1$ at
small $Q^2$ to $\approx 0.6$ at $Q^2\sim 10^3$ GeV$^2$. In the Regge
theory language this evidently implies that hard component of
neither ${\cal F}(x,Q^2)$ 
nor $G_{D}(x,Q^2)$ is dominated by a single Regge pole exchange and
a contribution from several hard Regge poles with broad spacing of 
intercepts is called upon. 
However, an approximately flat $Q^{2}$ dependence of $\Delta_{hard}(Q^2)$
shows that the hard component of the proton structure function can
well be approximated by a single Regge pole with intercept 
$\Delta_{hard} \approx 0.4$. Such a scenario in which a contribution of
subleading BFKL-Regge poles to $F_{2p}(x,Q^2)$ is suppressed 
dynamically because of
the nodal properties of gluon distributions for subleading BFKL-Regge
poles has been encountered earlier in the color dipole BFKL approach 
\cite{BFKLRegge}. The intercept $\Delta_{hard}(Q^2)$ found in the
present analysis is remarkably close the intercept of the leading
BFKL-Regge pole $\Delta_{\Pom}=0.4$ found in the color dipole 
approach in 1994 \cite{NZHERA,BFKLRegge,NZZ94}. For the related 
two-pomeron phenomenology of DIS see also \cite{LANDSH}.
From the point of view of $\bkappa$-factorization, the hard-to-soft
diffusion is a unique mechanism by which an approximate constancy of
$\Delta_{hard}(Q^2)$ derives from a very rapidly changing 
$\tau_{hard}(Q^2)$. Fourth, the same hard-to-soft diffusion provides
a mechanism for the rise of the real photoabsorption cross section
$\sigma^{\gamma p}$ in a model with the manifestly energy independent 
soft cross section. We emphasize that the hard-to-soft diffusion 
is a generic phenomenon and we do not see any possibility for 
decoupling of hard contribution from photoabsorption at $Q^2=0$.

We restricted ourselves to a purely phenomenological determination
of differential gluon distributions from the experimental data
on $F_{2p}(x,Q^2)$ which is sufficient for major applications
of the $\bkappa$-factorization technique. 
Whether the so-determined hard components 
of ${\cal F}(x,Q^2)$ and
$G_{D}(x,Q^2)$ do satisfy the dynamical evolution equations and
what is the onset of DGLAP regime will be addressed elsewhere. 

IPI wishes to thank Prof.~J.Speth for his hospitality
at Forschungszentrum J\"ulich. This work was partly supported by the grant
INTAS 97-30494.

\pagebreak

\end{document}